\documentclass[iop,author-year]{emulateapj}
\usepackage{url}
\usepackage{graphicx}
\usepackage{color}
\newcommand\etal{~et~al.}

\begin{document}
%%%%%%%%%%%%%%%%%%%%%%%%%%%
%    TITLE:               %
%%%%%%%%%%%%%%%%%%%%%%%%%%%
%\begin{portrait}

\title{The Spectacular Radio-Near-IR-X-ray Jet of 3C 111: X-ray Emission Mechanism and Jet Kinematics}
\shorttitle{The 3C 111 Jet}

\author{Devon Clautice\altaffilmark{1}, Eric S. Perlman\altaffilmark{1}, Markos Georganopoulos\altaffilmark{2}, Matthew L. Lister\altaffilmark{3}, Francesco Tombesi\altaffilmark{4,5}, Mihai Cara\altaffilmark{6}, Herman L. Marshall\altaffilmark{7}, Brandon Hogan\altaffilmark{3}, Demos Kazanas\altaffilmark{5}}

\altaffiltext{1}{Department of Physics and Space Sciences, Florida Institute of Technology, 150 W. University Blvd., Melbourne, FL 32901, USA}
\altaffiltext{2}{Department of Physics, University of Maryland -- Baltimore County, 1000 Hilltop Circle, Baltimore, MD  21250, USA}
\altaffiltext{3}{Department of Physics and Astronomy, Purdue University, 525 Northwestern Avenue, West Lafayette, IN 47907, USA}
\altaffiltext{4}{Department of Astronomy, University of Maryland, College Park, MD 20742-2421}
\altaffiltext{5}{NASA's Goddard Space Flight Center, Astrophysics Science Division, Code 663, Greenbelt, MD 20771, USA}
\altaffiltext{6}{Space Telescope Science Institute, 3700 San Martin Drive, Baltimore, MD 21218, USA}
\altaffiltext{7}{Kavli Institute for Astrophysics and Space Research, Massachusetts Institute of Technology, Cambridge, MA 02139, USA}
%\altaffiltext{5}{El Camino College NEED DETAIL}

\begin{abstract}

Relativistic jets are the most energetic manifestation of the active galactic nucleus (AGN) phenomenon.
AGN jets are observed from the radio through gamma-rays and carry copious amounts of matter and 
energy from the sub-parsec central regions out to the kiloparsec and often megaparsec scale galaxy 
and cluster environs.  While most spatially resolved jets are seen in the radio, an increasing number 
have been discovered to emit in the optical/near-IR and/or X-ray bands.  Here we discuss a 
spectacular example of this class, the 3C 111 jet, housed in one of the nearest, double-lobed FR II 
radio galaxies known.  We discuss new, deep {\sl Chandra} and {\sl HST} observations that reveal 
both near-IR and X-ray emission from several components of the 3C 111 jet, as well as both the 
northern and southern hotspots. Important differences are seen between the morphologies in the radio, X-ray, and near-IR bands. The long (over 100 kpc on each side), 
straight nature of this jet makes it an excellent prototype for future, deep 
observations, as it is one of the longest such features seen in the radio, near-IR/optical and X-ray 
bands. Several independent lines of evidence, including the X-ray and broadband spectral shape as well  as the implied velocity of the approaching hotspot, lead us to strongly disfavor the EC/CMB model and instead favor a two-component synchrotron model to explain the observed X-ray emission for several jet components. Future observations with  {\sl NuSTAR}, {\sl HST}, and {\sl Chandra} will allow us to further constrain the emission mechanisms.
 
\end{abstract}

\section{Introduction}

One of the milestone discoveries of {\sl Chandra} was the X-ray emission from nearly 100 quasar and
radio galaxy jets, as well as their hotspots{\footnote{see e.g., https://hea-www.harvard.edu/XJET/}}.  
The latter are high brightness 
regions where the jets collide with the intergalactic medium.  In the radio and optical, the 
emission from these sites is synchrotron in nature.  This
guarantees the presence of X-ray emission, via the 
Synchrotron Self Compton (SSC) process. The discrepancy between the observed X-ray fluxes
and the predictions of SSC models is often glaring
\citep[e.g.,][]{schwartz00,wilson01,samb04,marshall05}, with the X-rays commonly being orders of magnitude brighter than the SSC prediction if equipartition magnetic fields are assumed. 
\cite{tavecchio00} and \cite {celotti01} 
proposed to explain this excess X-ray emission as external Compton (EC) scattering of 
cosmic microwave background (CMB) photons by the jet's relativistic electrons. 
This requires jets with bulk Lorentz factor $\Gamma \sim 10$ that are oriented close to the line of 
sight for nearly their entire length.  
Alternatively \citep{dermer02}, the X-rays may be 
synchrotron emission from  high energy electrons suffering Compton losses
in the Klein-Nishina regime.  These particles are often required to be in a 
separate high-energy population \citep{hardcastle04,hardcastle06}.  In this case, the X-ray and optical
emission require {\it in situ} particle acceleration, as the radiating particles have lifetimes of a few to 
hundreds of years, much shorter than the particle's time to travel down the jet.  Those emissions 
would then provide an excellent probe of the physics in jet regions where particle acceleration is 
happening.  A third model \citep[upstream Compton,][]{georg03}, proposes a decelerating jet, 
with electrons in the faster, upstream flow 
scattering photons produced in the slow downstream flow, thus contributing to the X-ray emission.

Discriminating between these models relies on several diagnostics, including component 
spectral energy distributions (SEDs) and differences between radio and X-ray jet morphology
\citep{Jester06,Kharb12}).  
We have proposed two diagnostics that can rule out the EC/CMB model.
The first of these %\citep{Uchiyama:11,
\citep{Krawczynski:11, Poutanen:93} relies on the fact that  except for scatterings from 
low-energy particles ($\gamma \sim 1$) inverse-Comptonized CMB 
radiation should be unpolarized, reflecting the unpolarized nature of the seed photon population.
This diagnostic was first used by \cite{cara13} to almost completely rule out the EC/CMB model for one 
quasar jet, PKS 1136-135. 
Another diagnostic \citep{georg06} relies on the fact that the observed synchrotron emission at 
IR and lower energies must also be Comptonized, resulting in a minimum level of GeV gamma-ray
emission.  This has ruled out the EC/CMB model for the jets of 3C 273 and PKS 0637-752
\citep{meyer14,meyer15}).  Finally, in a few FR IIs (e.g., Pictor A, \cite{hardcastle15,gentry15,marshall10}) that are viewed at larger angles, the broadband SED even 
suggests synchrotron emission without requiring a separate, high-energy electron population. 

\begin{figure*}[t]
\begin{center}
%	\vspace{-18pt}
	%\captionsetup{width=1.00\textwidth}
%		\begin{minipage}{0.4\textwidth}
			\includegraphics [keepaspectratio=true,width=10.7cm]{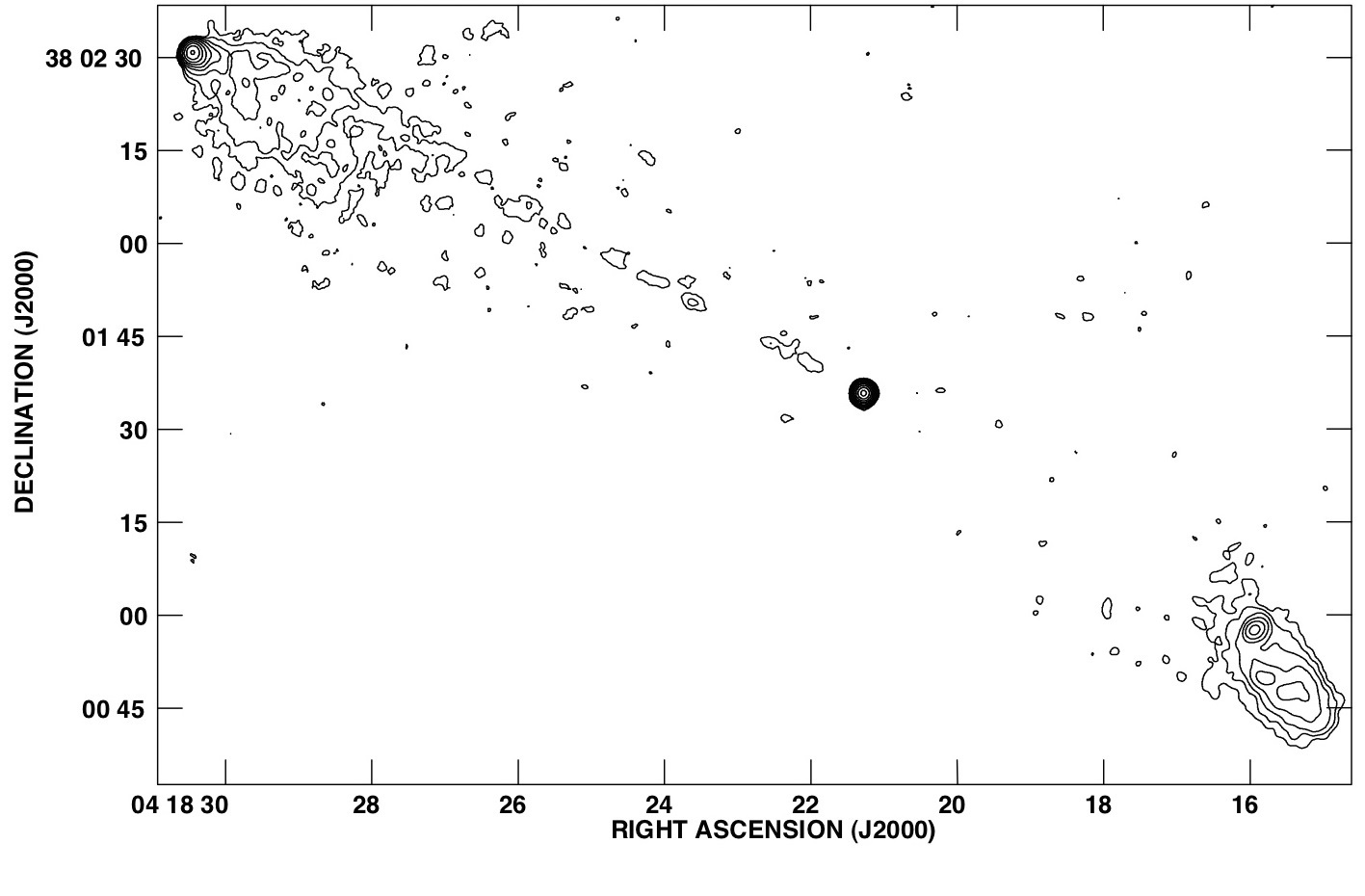}
			\includegraphics [keepaspectratio=true,width=6.7cm]{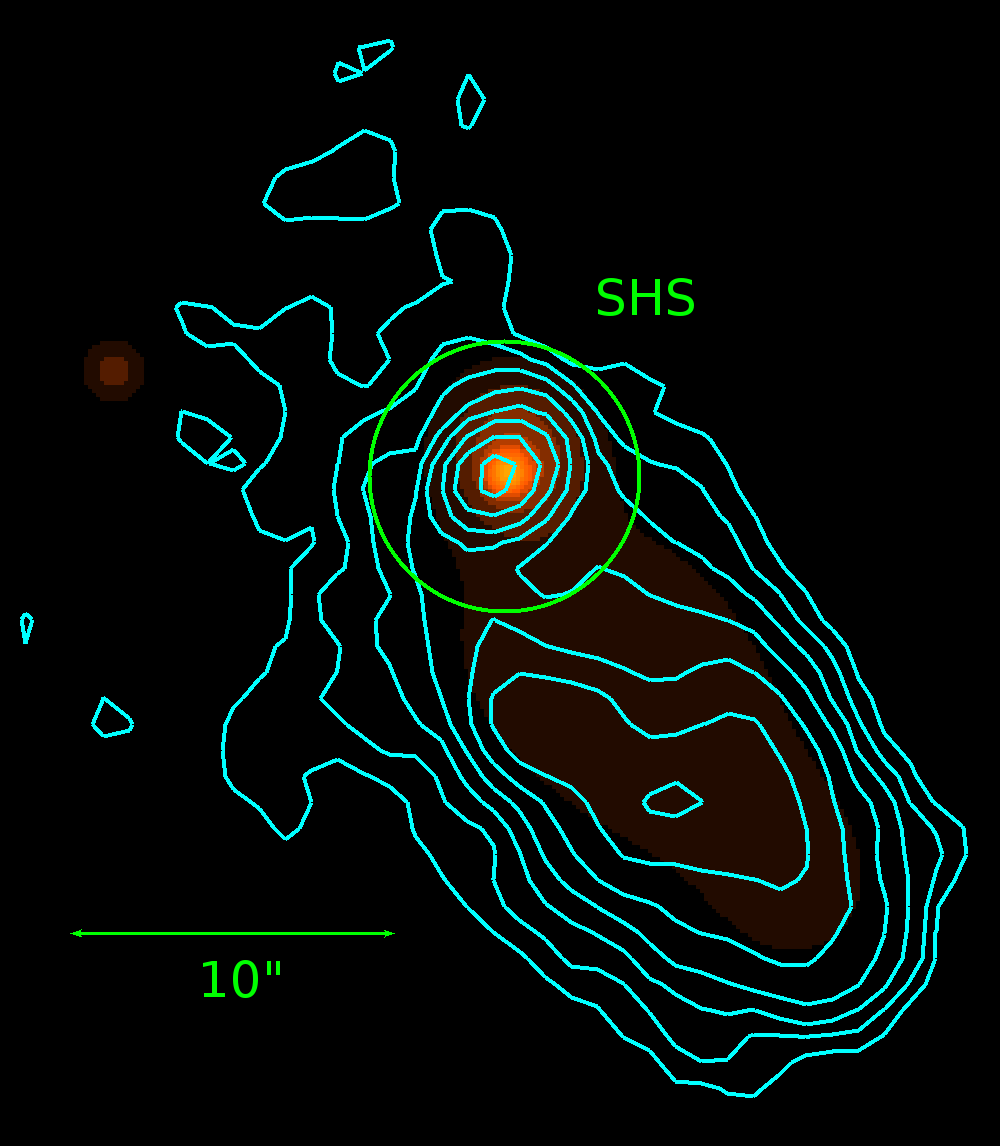}

%			\end{minipage}
%		\begin{minipage}{0.48\textwidth}
			\includegraphics[keepaspectratio=true,width=17.5cm]{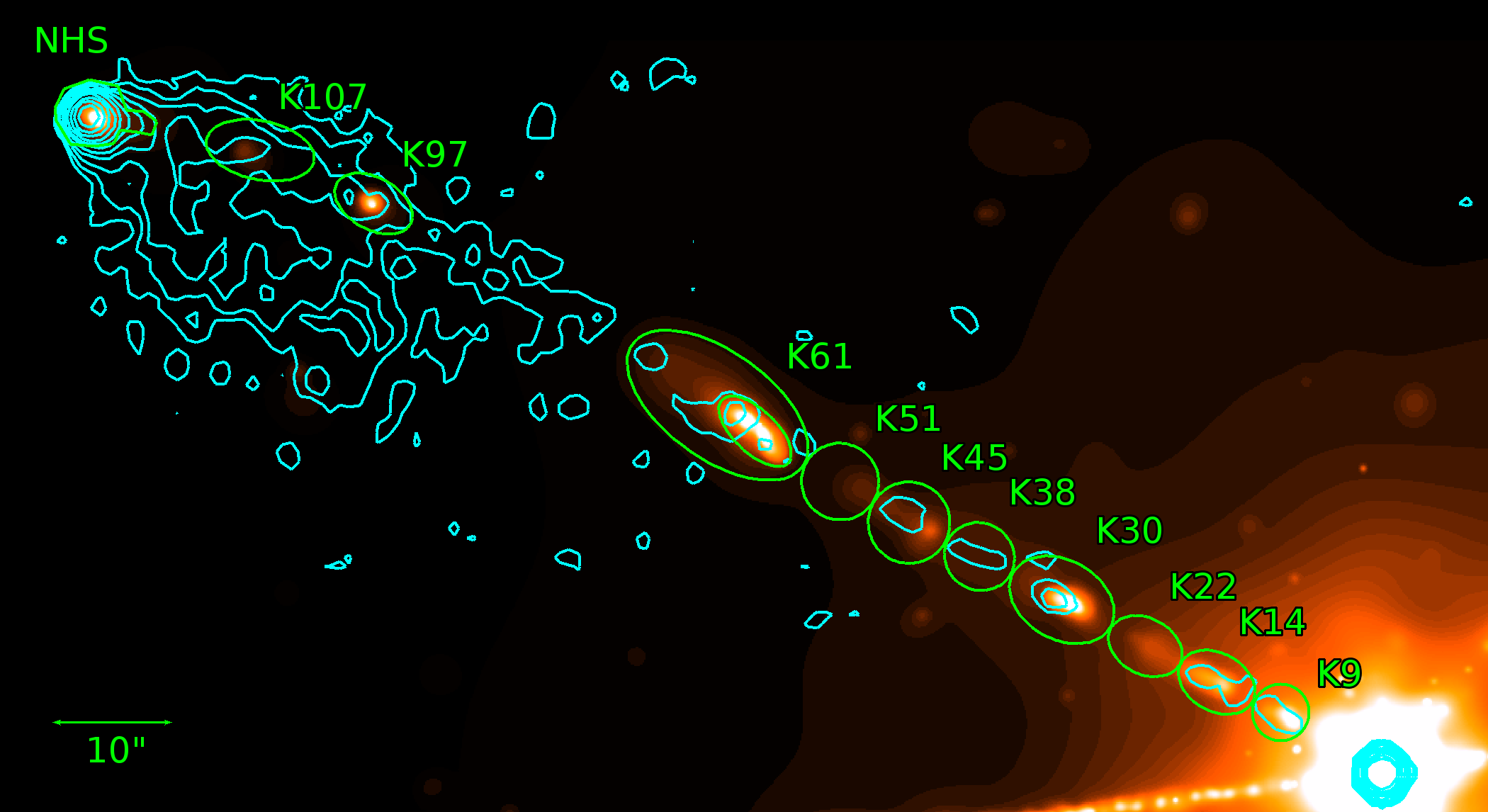}
%			\includegraphics[keepaspectratio=true,width=3.4in]{3c111_xray_with_regions.png}	%			\end{minipage}
%		\begin{minipage}{0.1\textwidth}
%		\end{minipage}
\end{center}
		\caption{ 3C 111, as seen in a 1989 8 GHz VLA observation \citep[top left,][]{leahy97}, and a deep {\sl Chandra} X-ray observation.  The {\sl Chandra} image, shown at top right and bottom, shows the X-ray image displayed with a heat scale and radio emission overlaid as cyan contours.  
%		{\sl Note the extraordinary length of the jet:  over 2 arcminutes. }  
The {\sl
Chandra} image shows emission from eight jet regions (named in green), as well as the 
northern and southern hotspots (NHS and SHS). Also shown in green are the flux extraction regions for each knot region.  Note that significant differences exist between
the morphology seen in the two bands.  See \S\S 2, 3 for discussion.}

\end{figure*}

%While spectral energy distributions (SEDs) 
%are helpful, multiple emission models can almost always reproduce a given SED, even when 
%there is multiwavelength data.  Imaging by itself is also not adequate, as simple superposition 
%is very difficult to deal with.  Polarimetry can make a better discriminant, as EC-CMB emission 
%is unpolarized except at the very lowest particle energies, due to the unpolarized nature of the seed 
%photon population, but it requires either a bright jet or very deep observations due to the need for 
%extremely high S/N at reduced efficiency.  
With all of these different possibilities, one of the most 
basic needs for investigating models of both jet emission and physics is to find ideal testing grounds.  Only 
a very few prototype jets, that are bright in several bands at low redshifts, and minimally bent, are
known. 
%
%Only a
%limited number of real prototype objects are known:  these are objects that are bright in several bands,
%at low redshift (so that a high resolution can be attained in every waveband), and as long
%as possible (so that a wide variety of features can be examined in detail).  Also it is preferable if the 
%jet is straight, so that there is minimal effect from bending, which not only introduces considerable
%hydrodynamical complications but also will change completely the observed spectrum of a jet, 
%since the observed luminosity is proportional to a high power of the  
%beaming factor $\delta = [\Gamma (1-\beta \cos \theta)]^{-1}$, where $\Gamma$ is the bulk Lorentz
%factor, $\beta=v/c$ and $\theta$ is the jet's viewing angle.
Here we discuss a new, prototype jet.
%that has the distinction of being the 
%lowest-redshift member of its class where emission is seen in the radio, near-IR and X-ray bands.  
%By virtue of its low redshift, it is also the longest such object on the sky.  

3C 111 is a powerful FR II radio galaxy \citep{FR:74} at 
$z=0.0485$ \citep{HB91}.  Our {\it HST} images (\S\S 2-3) show that the host galaxy is a bright giant elliptical with somewhat distorted outer isophotes, and 
several prominent  companions within 50 kpc.  On parsec scales, VLBI observations show 
component speeds as high as 8$c$ in the approaching, northern jet \citep{lister13}. 
Shallow, 10 ks {\sl Chandra X-ray Observatory} survey observations  by 
\cite{hogan10} revealed X-ray emission from three knots in the northern jet (which we call K30, K61 and K97) and the 
northern hotspot (NHS).  The jet is extremely long (nearly 4 arcminutes) and its host galaxy 
resides in a rich optical environment.  Here we 
discuss the results of new, deep observations with both {\sl Chandra} and the {\sl Hubble Space 
Telescope} 
(HST).  These observations not only confirm the results of \cite{hogan10} but also 
reveal near-IR and X-ray emission from several components in the 3C 111 jet, as well as the 
southern (receding) hotspot.

This paper is laid out as follows:  In Section 2, we describe our observations  and data reduction
methods.  Section 3 shows the results and discusses the broadband spectrum of the jet components.
We close in Section 4 by stating our conclusions.  
Throughout this paper we assume $\Omega_m=0.27$,
$\Omega_{\Lambda}=0.73$, $\Omega_r=0$ and $H_0=71\mbox{ km}\mbox{
s}^{-1}\mbox{ Mpc}^{-1}$.  
%We also adopt the following convention for the
%spectral index, $\alpha$: $F_{\nu}\propto\nu^{-\alpha}$, where $\nu$ is the
%frequency.   

\begin{deluxetable*}{ccccccc}
\tablecolumns{7}
\tablewidth{0pt}
\tablecaption{{\sl HST} Observations of 3C 111}
\tablehead{\colhead{Date} ~~~& \colhead{Program} & \colhead{Instrument} & \colhead{Band} & 
\colhead{$\lambda_{pivot}$ (nm) } & \colhead{ width (nm)} & \colhead{$T_{int} (s)$}}
\startdata
30/01/2013 & 13114 & WFC3/UVIS & F850LP & 916.6 & 118.2 & 2534 \\
30/01/2013 & 13114 & WFC3/IR & F160W & 1536.9 & 268.3 & 2606$^a$ \\
26/02/1996 & 5931 & WFPC2 & F791W & 788.1 & 123.1 & 53400$^b$ \\
08/12/2004 & 10173 & NICMOS/NIC2 & F160W & 1600 & 400 & 1152$^c$ \\
19/11/1995 & 5476 & WFPC2 & F702W & 691.9 & 138.5 & 600$^d$ 
\enddata
\tablenotetext{a}{Two pointings. Integration time is per pointing.}
\tablenotetext{b}{Field does not include host galaxy or optically seen part of jet.}
\tablenotetext{c}{Field includes only one knot region (K9) and image is not deep enough to 
confirm its detection.}
\tablenotetext{d}{Very shallow image, does not show jet, not used.}

\end{deluxetable*}

\section{Observations and Data Reductions}

\subsection{{\sl Chandra} Observations}

{\sl Chandra} has observed 3C 111 three times with the ACIS-S.  In 2008, a shallow, 10 ks survey observation (dataset 701719)
was taken \citep{hogan10}, which discovered the X-ray emission from the jet.  
On 10-11 January 2013, we obtained much deeper observations (dataset 702798), for a total on-source time of 127 ks.  These observations were gathered using alternating exposure mode, 
with interleaved frame times of 1.5s ($\times 4$) and
0.3s ($\times 1$) during each cycle.  This was done in order to enable us to minimize the effect of 
pileup in the region of the quasar nucleus, while at the same time keeping the majority of the time 
optimized for detection of fainter sources in a broader field.  It 
reduced efficiency by 15\%, giving us a total exposure time of
92 ks (1.5s frame time only), but allowed us to discriminate inner jet knots from emission due to the AGN in the innermost 
10 arcseconds, where pileup is a factor, by using the 0.3s frame time data (17 ks exposure time).   These observations were augmented on 4-5 November 
2014 by 
ACIS/HETG observations (150 ks, PI F. Tombesi, dataset 703007).

 All 
observations were reduced in
CIAO version 4.8.0, using CALDB v. 4.7.0, with standard screening criteria and calibration files provided by the {\sl Chandra} X-Ray Center. Pixel randomization was removed, and only events in grades 0, 2 Ð- 4, and 6 were retained. We also checked for flaring background events. No significant flaring events were found, so that we did not have to filter by time.  We subsampled the native {\sl Chandra} resolution by 4, leading to a pixel scale of 0.123 arcsec/pixel.
%Figure 1 shows the {\it Chandra} observations along with archival {\it VLA} radio observations (ref), showing that X-ray emission can be seen from eight different regions in the approaching northern jet, as well as the northern hotspot.  
Datasets 702798 and 703007 were combined to obtain the images discussed in this paper. We chose not to incorporate the much shallower dataset 701719 into that analysis because of its poor statistics.
To show the extended structure, we smoothed the X-ray image
adaptively using the CIAO task {\it csmooth}, smoothing only below a minimum significance of 4.
{\footnote{This smoothed image was not used for scientific measurements, but is useful for illustrative purposes.}}

\subsection{{\sl HST} Observations}

\begin{figure*}[t]
\includegraphics[width=8.6cm]{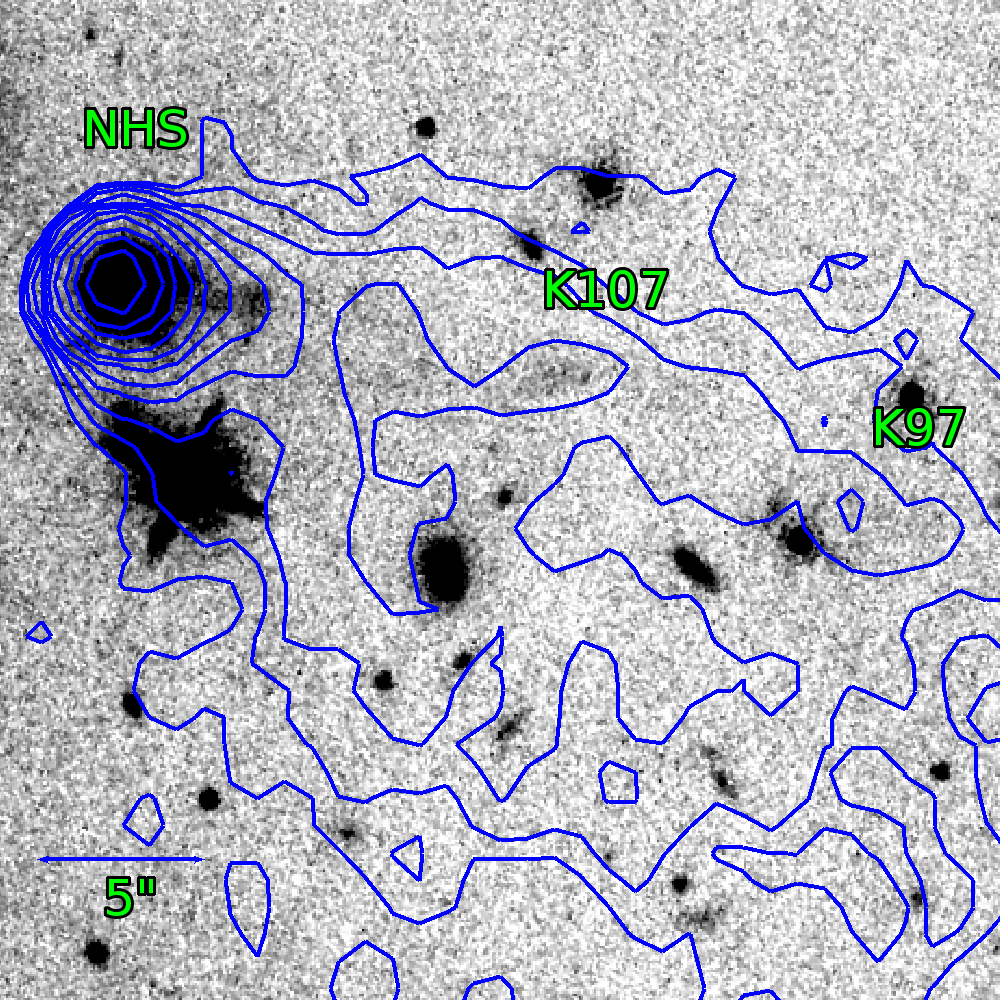}~~
\includegraphics[width=8.6cm]{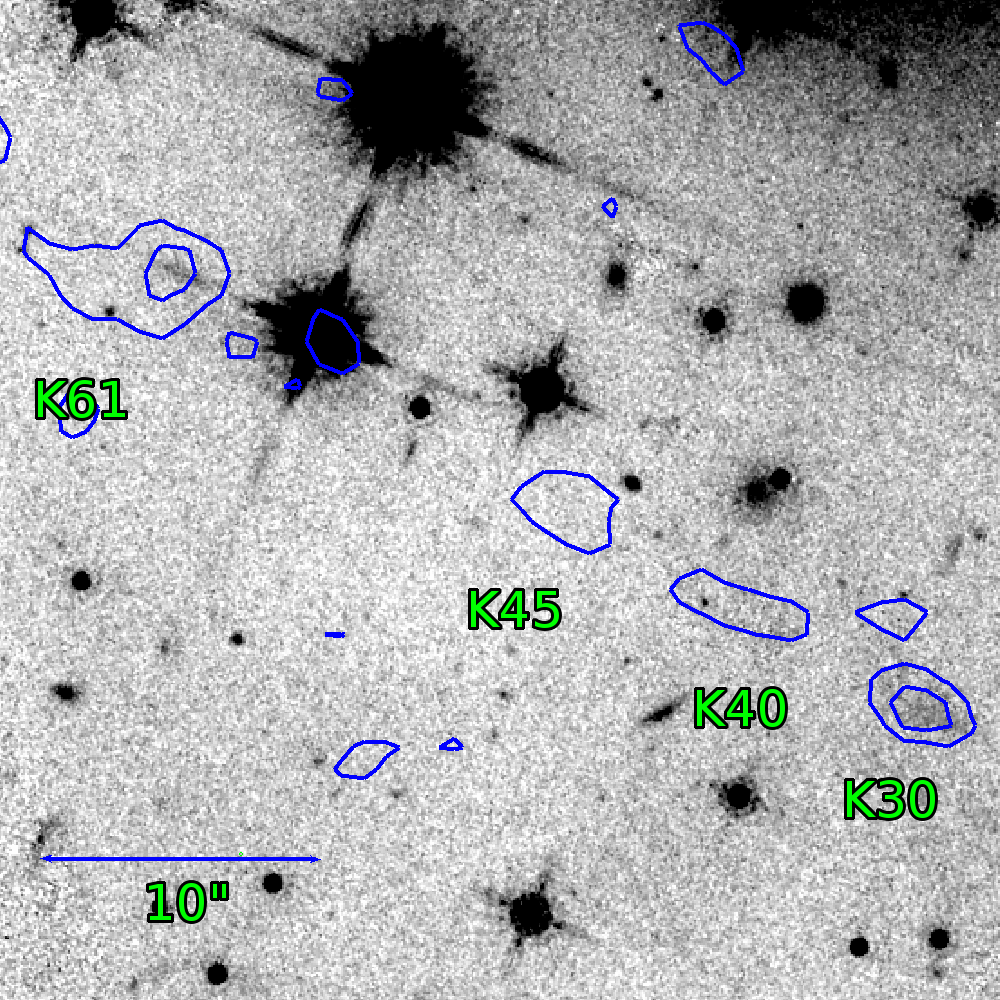}
\includegraphics[width=8.6cm]{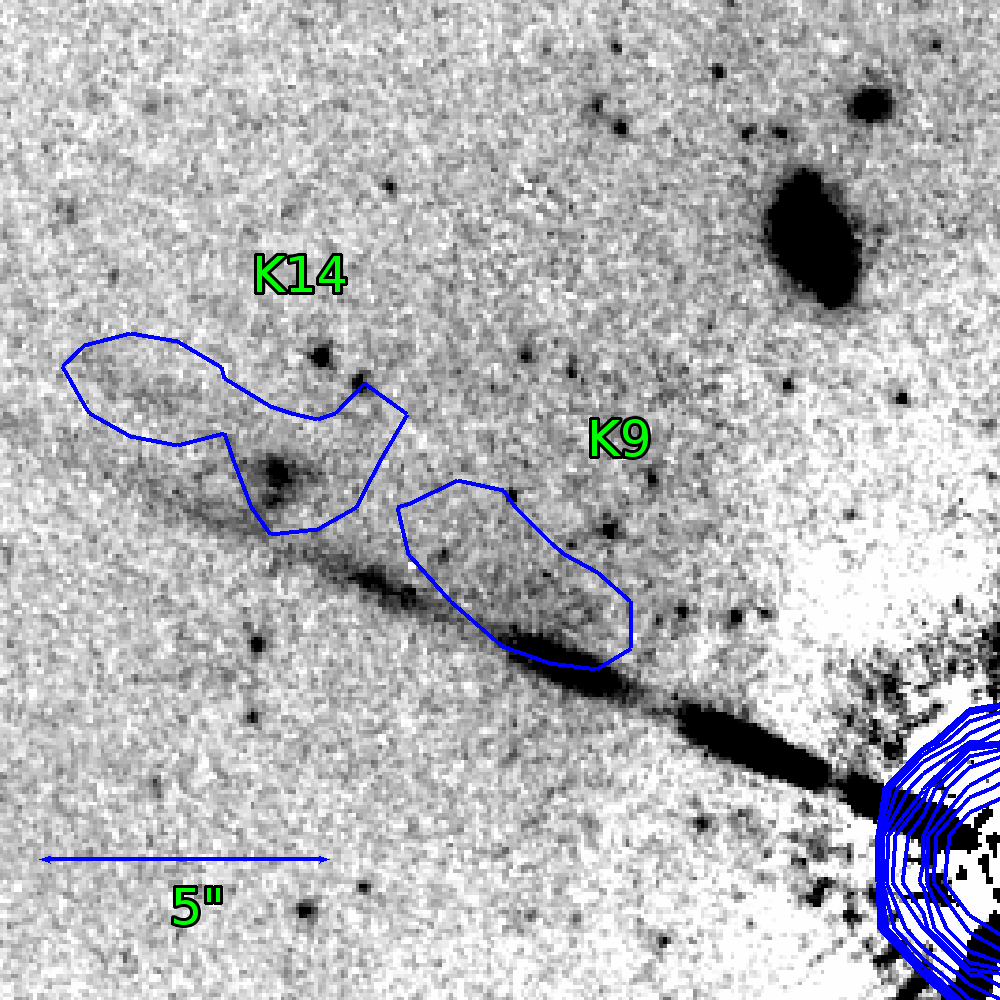}~~~~~~~~
\includegraphics[width=8.6cm]{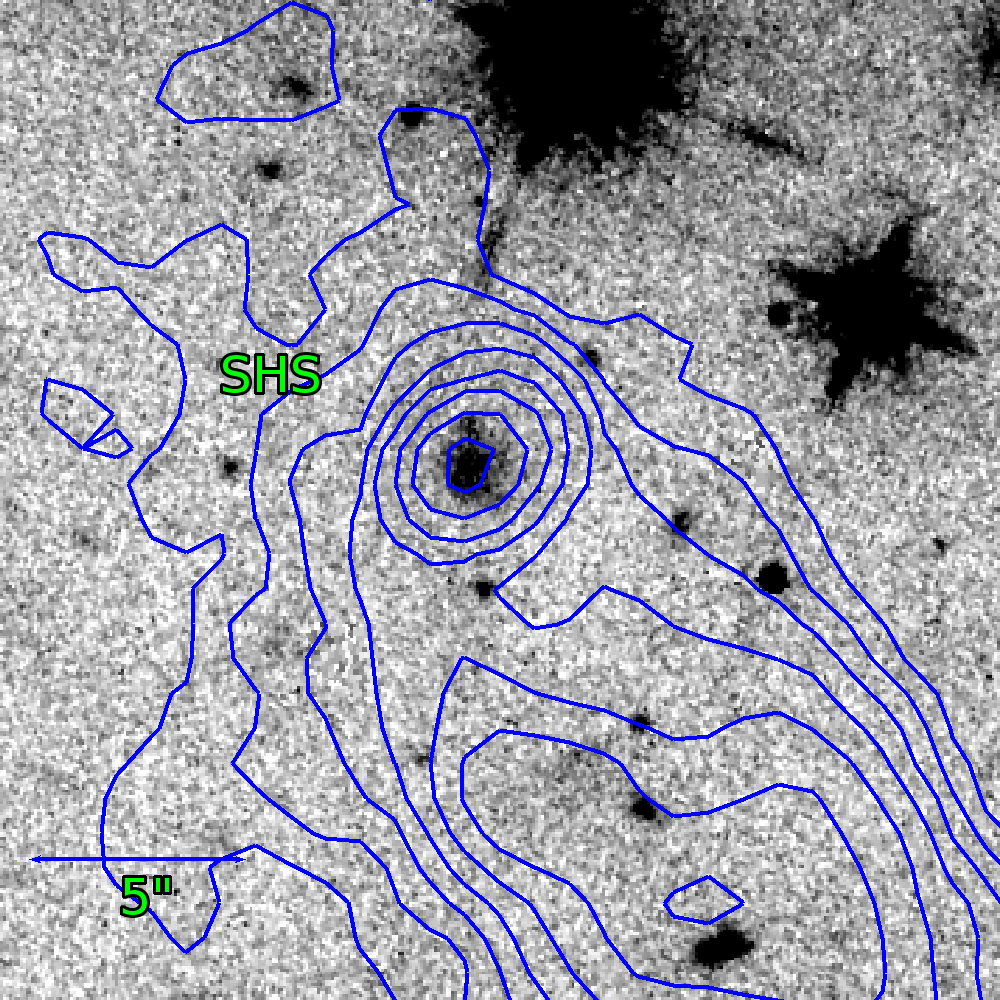}

\caption{{\sl HST} images of the jet of 3C 111, obtained with the WFC3/IR + F160W.  All four panels show 
the HST image in greyscale, with contours from the radio image over-plotted in blue.   A galaxy model has been subtracted from the HST image (see \S 3).  At top left, we show the 
brighter, northern hotspot, which is well resolved both parallel to and perpendicular to the jet direction. 
The top right and bottom left images show the inner part of the northern jet, 
specifically the part extending from about 
$5\arcsec$ to about $65\arcsec$ from the nucleus.    The bright linear feature in the bottom left panel is a diffraction spike. At bottom
right, we show the fainter, southern hotspot, which is also well resolved. The X-ray detected knots have been labelled in green.}
\end{figure*}

{\sl HST} observed 3C 111 on 30 January 2013 for three orbits, using the Wide-field Camera 3 
(WFC3).  Images were gathered both in the UVIS channel using the F850LP filter (1 orbit) and in the 
IR channel
using the F160W filter (2 orbits).  Because of the size of the 3C 111 jet-hotspots system, 
we restricted  {\it HST}'s orientation so that the jet fell along a chip diagonal in both observations.  
Unfortunately for ease of scheduling we had to leave a 10-degree allowance on the allowed position
angle (PA), and the PA that was used placed 
%Unfortunately because of the sheer size of the possible emission regions in both hotspots (particularly
%the southern one) it was very difficult to pinpoint the correct point position and as a result, 
emission 
from the NHS at the edge of the field of the UVIS/F850LP observation.
To compensate for this, we located archival observations obtained on 26 February 1996 with 
the  Wide-Field and Planetary Camera 2 (WFPC 2) with the F791W filter (PI Meisenheimer).  These 
latter observations include only the northern hotspot and some of the northern jet.
Because of the small field of view, two pointings were necessary in the IR channel, while 
one pointing was deemed adequate in the UVIS.  In addition, we used a standard, 2-position 
dither pattern at each location in each band.  This, combined with the multiple readouts, was more
than adequate to remove bad pixels and cosmic rays in the IR/F160W observation, but in the UVIS/
F850LP observation it was not adequate, and there were a significant number of pixels that had 
cosmic ray strikes in both images. 
In addition to the above there exist two shorter observations obtained with NICMOS and 
WFC2 (PI Sparks).  Table 1 gives details of all {\sl HST} observations.  

All {\sl HST} images were re-calibrated in PYRAF using the most up-to-date reference files (i.e., flat field, distortion 
correction table, etc.) obtained from the STScI Calibration Database system.  We corrected for charge 
transfer efficiency (CTE) effects in the UVIS/F850LP data using the recipe of \cite{And2012} and in
the WFPC2 data using the recipe of \cite{Dolphin:00} and \cite {Riess:00}.  In the UVIS/F850LP data we also pre-processed the data 
using L.A. Cosmic\footnote{see http://www.astro.yale.edu/dokkum/lacosmic/} \citep{vanDok:01} 
prior to drizzling.  This significantly decreased the number of cosmic rays affecting the final image.
We used the  \verb|Astrodrizzle| task \citep{Gonzaga:12} from the
\verb|STSCI_PYTHON| package to drizzle-combine the images for each of the two filter combinations.
Besides combining the images, \verb|Astrodrizzle| distortion-corrects the images, performs image flat-fielding, cosmic-ray
rejection, image alignment, and other tasks.    
Prior to any analysis, the HST data had to be galaxy-subtracted.
This was done using the tasks \verb|ellipse| and \verb|bmodel|. 
%A constant value of 0.4138 counts/pixel was also subtracted to
%represent sky emission.  
The model fitting was done iteratively, excluding nearby stars and galaxies (note that 3C 111 lies in a fairly dense cluster of galaxies).  
%Details of this model will be given in a 
%followup paper.

Local background regions were used to determine the blank sky noise emission
for each source aperture. Sigma clipping was used to eliminate any pixel values that deviated beyond 
3 sigma from the median. Photon noise was estimated by multiplying the weight map created by 
\verb|Astrodrizzle| with our science image (in counts/second) to obtain the number of counts in each sky-
subtracted source region. Read noise was taken from the header values in each image; dark current 
was estimated from the dark reference file indicated in the header.

Aperture photometry was done on the images using the apertures shown in Figure 1. Aperture 
correction was done following the recommendations of the WFC3 Data Handbook \citep{rajan11}, 
while for the WFPC2 dataset it was done following  \cite{holtzmann95}.
Conversion to flux units was performed by multiplying image data in electrons/s by the corresponding \hyphenation{PHOTFLAM} and \hyphenation{PHOTPLAM} values for all images.
3C 111 is at a low galactic latitude ($b_{II}=8.8^\circ$), relatively near the 
Taurus molecular cloud (the nearest large star-forming region in our Galaxy). 
Ungerer et al. (1985), in their detailed optical and radio study, pointed out that the region of the cloud in front of 3C 111 is not the densest part (see Figure 3 of Ungerer et al. 1985). This result is also supported by the results of the XMM-Newton Extended Survey of the Taurus Molecular Cloud project (G\"udel et al. 2007). Due to the presence of this molecular cloud, galactic extinction 
is unfortunately high, with $A_V=4.5$ mag assuming a standard $R_V=3.1$ \citep{SF11,SFD98}.  
We note that \cite{meisenheimer97} used a much lower value for the extinction to 3C 111,
stemming from the earlier survey of \cite{burheil82}
(see also \S 3).

\section {Results}

The 3C 111 jet can be seen across the electromagnetic spectrum, from the radio through the X-rays.
In Figure 1, we show our deep {\sl Chandra} imaging of 3C 111, along with archival VLA imaging 
\citep{leahy97}.  X-ray emission is evident in at least 8 jet regions, plus the northern and 
southern hotspot.  This emission is also seen in the near-IR, as shown in Figure 2, which shows 
close-ups of three jet regions in the F160W image, respectively the northern hotspot, inner jet and 
southern hotspots.  The near-IR image shows emission from most, but not all X-ray emitting jet 
regions.
%, and also shows evidence of differences between the radio, near-IR and X-ray morphology.  
In the F850LP and F791W images, the only jet or jet-related emission that can be seen comes from 
the northern hotspot.
%, although when the F850LP and F791 images are co-added, one knot region 
%(K61) shows up at about 4 $\sigma$ (TRUE?).   
This is likely a result of the high Galactic extinction towards 3C 111.  
Most of the panels in Figures 1 and 2 show one image as greyscale and another as contours, 
allowing us to
compare the morphology in different bands.  
To aid in this comparison, we named the northern jet features using 
the distance in arcseconds from the nucleus.  Thus, as an example, knot K14 has its flux maximum 14\arcsec~ from the nucleus.  

\subsection{Jet Morphology}

There is significant evidence of differences between the radio, 
near-IR and X-ray morphology, as seen in Figures 1 and 2, as well as in Figure 3, which shows the 
profile of relative flux (each normalized to 1 at an arbitrary point) along the jet in the 
{\it Chandra}, F160W, 
and VLA images.   We note that there are strong differences between the radio, near-IR and X-ray 
fluxes. The near-IR and X-ray morphology are discussed in detail here. The radio morphology will 
be discussed in more detail in a future paper, where we also discuss follow-on deep JVLA
observations.  In the next sub-section, we will discuss the spectral energy distribution of the jet 
features, including the X-ray and optical spectral indices for the knots where it was possible to 
extract such information.  In registering the three data sets to a common frame of reference, we assumed the VLA map to be the fiducial, adhering to the usual IAU standard. The {\sl HST} images
were registered to this frame by hand, as the Guide Star Catalog alignment always has errors of 
near arcsecond level, assuming that the optical and radio AGN core positions were identical.  To register the 
{\it Chandra} data to this frame, we followed the CIAO thread ``Correcting Absolute Astrometry", using 
CIAO task {\sl wavdetect} to match sources in the 2MASS catalog in the {\it Chandra} images.  This 
yielded a final offset of about 0.2\arcsec~ from the radio.  We also merged the data from datasets 702798
and 703007 using {\sl reproject\_obs} in CIAO.  Following this, the $1\sigma $ errors in the positions from the {\it HST} image are $<0.02\arcsec$, while those in the X-ray image are 
$\pm 0.16\arcsec$ relative to the radio frame of reference according to \cite{Rots11}, although to be conservative for this purpose we used 0.3\arcsec.

\begin{figure}
\includegraphics[width=9.5cm]{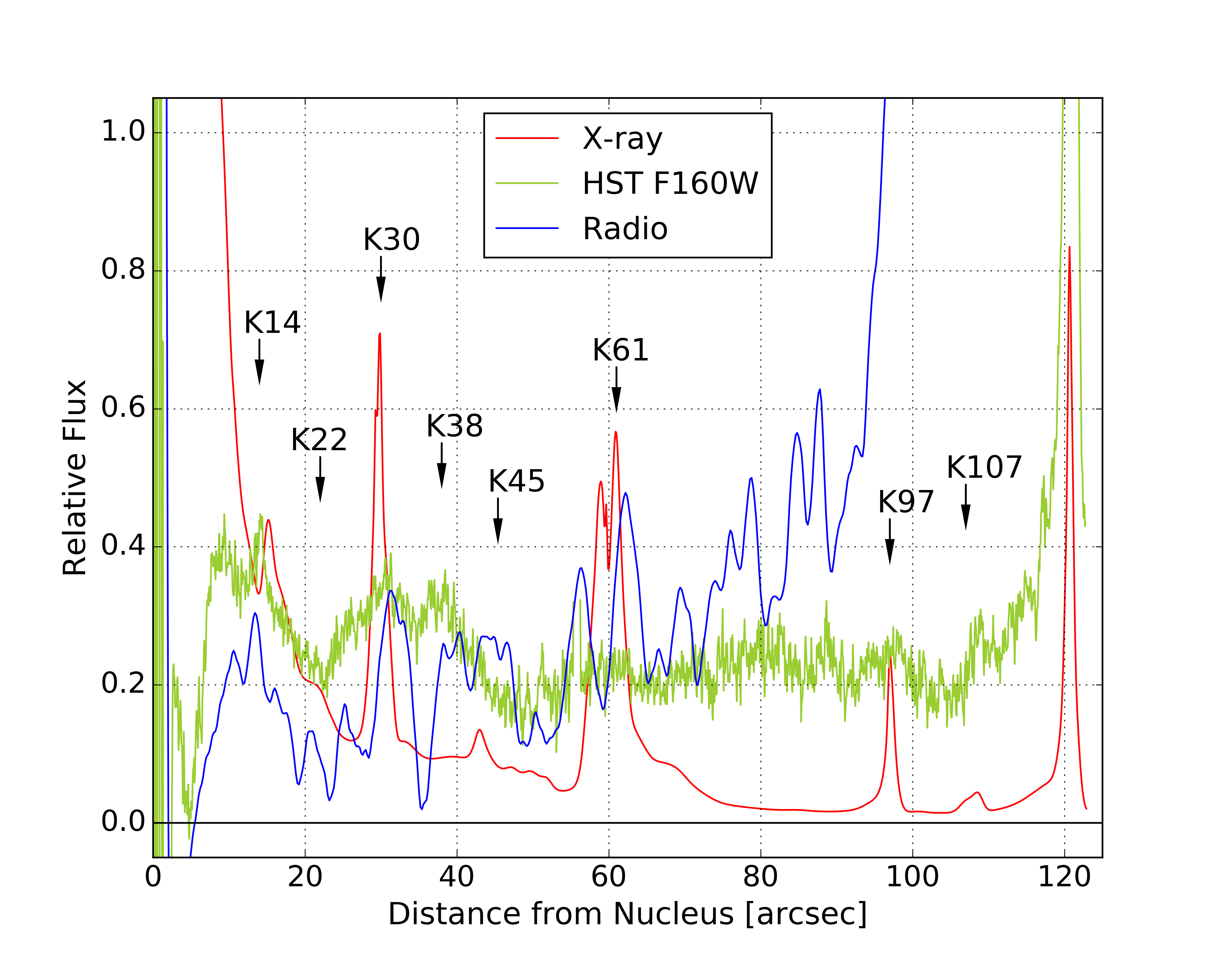}
\caption{Relative flux as a function of distance from the nucleus for the approaching 
jet of 3C 111, as seen in the radio ({\it VLA} 
image, blue trace), near-IR ({\it HST/WFC3 IR/F160W} image, green trace) and X-rays ({\it Chandra} 
image, red trace).  Each of the three traces was extracted from a slice $1.476\arcsec$ wide, along a vector  extending from the nucleus of the galaxy through the NHS.   Major knot 
regions are labeled above the traces.  See \S 2 for
details on the alignment of the three images, and see \S 3 for discussion.}
\end{figure}

Components within $\sim$ 20\arcsec~ of the nucleus have flux profiles in 
the X-rays that are mixed with that of the unresolved nuclear source due to the {\sl Chandra} PSF (see Fig. 3), and are somewhat piled up in the long frame time, undispersed {\sl Chandra} image, 
and within 10-15\arcsec~ the knots also lie within the galaxy seen in the optical/near-IR image.  
However, despite this, we can make a few remarks about how the radio, optical and X-ray 
morphologies compare, using the short frame time data from the interleaved dataset (702798) 
as well as the Order 0 HETG image (dataset 703007).  Knot K9's X-ray 
morphology (Figure 1) has a strong peak towards its upstream end 
that is not seen in the radio.  Unfortunately, however, it lies too close to the 
diffraction spike in the F160W image to fully characterize in the near-IR. Knot
K14 appears to peak slightly closer to the nucleus in the radio image than in either the near-IR or X-
rays.  X-ray emission is clearly seen downstream of that component extending nearly continuously to 
knot K30.  That emission is not seen in either the near-IR or the radio (the near-IR emission that is
present is more likely due to subtle galaxy features in the same region, as shown in the middle panel 
of Figure 2).  That X-ray emission includes a knot seen only in the X-rays, knot K22. 

\begin{deluxetable*}{crrlll}
\tablecolumns{6}
\tabletypesize{\small}
\tablewidth{0pt}
\tablecaption{Jet Component Flux Densities}
\tablehead{\colhead{Region} & \colhead{Radio} &\colhead{F160W} & \colhead{F850LP} &\colhead{F791W} & \colhead{2 keV} \\
& \colhead{mJy} & \colhead{$\mu$Jy}& \colhead{$\mu$Jy}& \colhead{$\mu$Jy}& \colhead{nJy}} \\
\startdata
K9 & $6.91 \pm 0.30$ & $4.88 \pm 0.12$ & $< 8^a$ &    ...$^e$  & $1.07 \pm 0.20$\\
K14 & $9.35 \pm 0.44$ & $5.38 \pm 0.15$ & $< 10^a$ &   ...$^e$  & $0.91 \pm 0.16$\\
K22 & $3.78 \pm 0.42$ & $0.46 \pm 0.20$  & $<11^a$ &  ...$^e$ & $0.40 \pm 0.11$\\
K30 & $13.15 \pm 0.60$ & $8.75 \pm 0.23$ & $...^{a,b}$ & $<11^a$ &   $2.26 \pm 0.26$    \\  
K40 & $8.25 \pm 0.44$ & $2.13 \pm 0.20$ & $<14^a$ &   $<7^a$ &   $0.35 \pm 0.09$\\
K45 & $11.96 \pm 0.52$ & $1.13 \pm 0.25$ & $<17^a$ &  $<9^a$ &    $0.55 \pm 0.11$\\
K51 & $6.03 \pm 0.49$ &$1.72 \pm 0.18$ & $<16^a$ & $<8^a$ &  $0.22 \pm 0.08$\\
K61 & $ 38.13 \pm  0.95$ & $1.77 \pm 0.46^{b}$ & $<22^{a,c}$ &$<23^{a,c}$ & $4.38 \pm 0.40$\\
K97 & $ 27.88 \pm 0.42 $ & $2.01 \pm 0.28$ & $<11^a$ & $<6^a$ & $0.59 \pm  0.09$\\
K107 & $ 58.00 \pm 0.51 $ & $4.70 \pm 0.36$ & $<16^a$ & $<12^a$ & $0.18 \pm 0.06$\\
NHS & $ 610.31 \pm 0.47 $ & $121.0 \pm 0.6$ &...$^d$ & $44.3 \pm 4.0$ & $1.86 \pm 0.21$\\
SHS & $ 182.52 \pm 0.63 $ & $13.39 \pm 0.53$ & $<22^a$ & ...$^e$ & $0.35 \pm 0.09$
\enddata
\tablenotetext{a}{Flux quoted is a 2$\sigma$ upper limit.}

\tablenotetext{b}{Located in chip gap.}
\tablenotetext{c}{Bright star plus diffraction spikes within region.}
\tablenotetext{d}{At chip border, significant part of region off chip.}
\tablenotetext{e}{Knot is off chip.}
\end{deluxetable*}

Knot K30, seen 
in all three bands, has an X-ray flux peak that is located significantly upstream of either the radio or near-IR one (Fig 4), with the near-IR peak located closer to the nucleus than the radio one. 
The X-ray flux from K30 also declines much more quickly with increasing distance from 
the nucleus than in either the near-IR or radio, which show similar decline rates (Figure 3).  The 
K40-K45 region is also complex.  The radio flux of K40 displays two peaks, with the near-IR 
peak associated with the one closer to the nucleus.  The X-ray emission, however, does not peak 
until 42\arcsec~ from the nucleus, where there is an apparent radio minimum.  The radio emission picks up 
again between 43-45\arcsec, while through that region and extending out to nearly 50\arcsec, the X-ray emission 
appears roughly continuous and there is no significant X-ray knot at 51\arcsec~ from the nucleus as there is in the radio.  
Moving further out, there is a flux maximum at about 55\arcsec~ from the nucleus in the radio 
image that is not seen in the X-ray or F160W images.  Knot K61, which represents an apparent 
'kink' in the jet, is seen in both the radio and X-rays. Its X-ray morphology has a 'corkscrew' like 
appearance that is not prominent in the radio, where only its downstream half can be seen.  
In the optical/near-IR, K61 unfortunately lies very near a bright foreground 
star and so while there is possible
emission in the near-IR it lies too close to the star or its diffraction spikes to have confidence in its 
detection and/or measure a flux.

\begin{figure}

\includegraphics[width=9.cm]{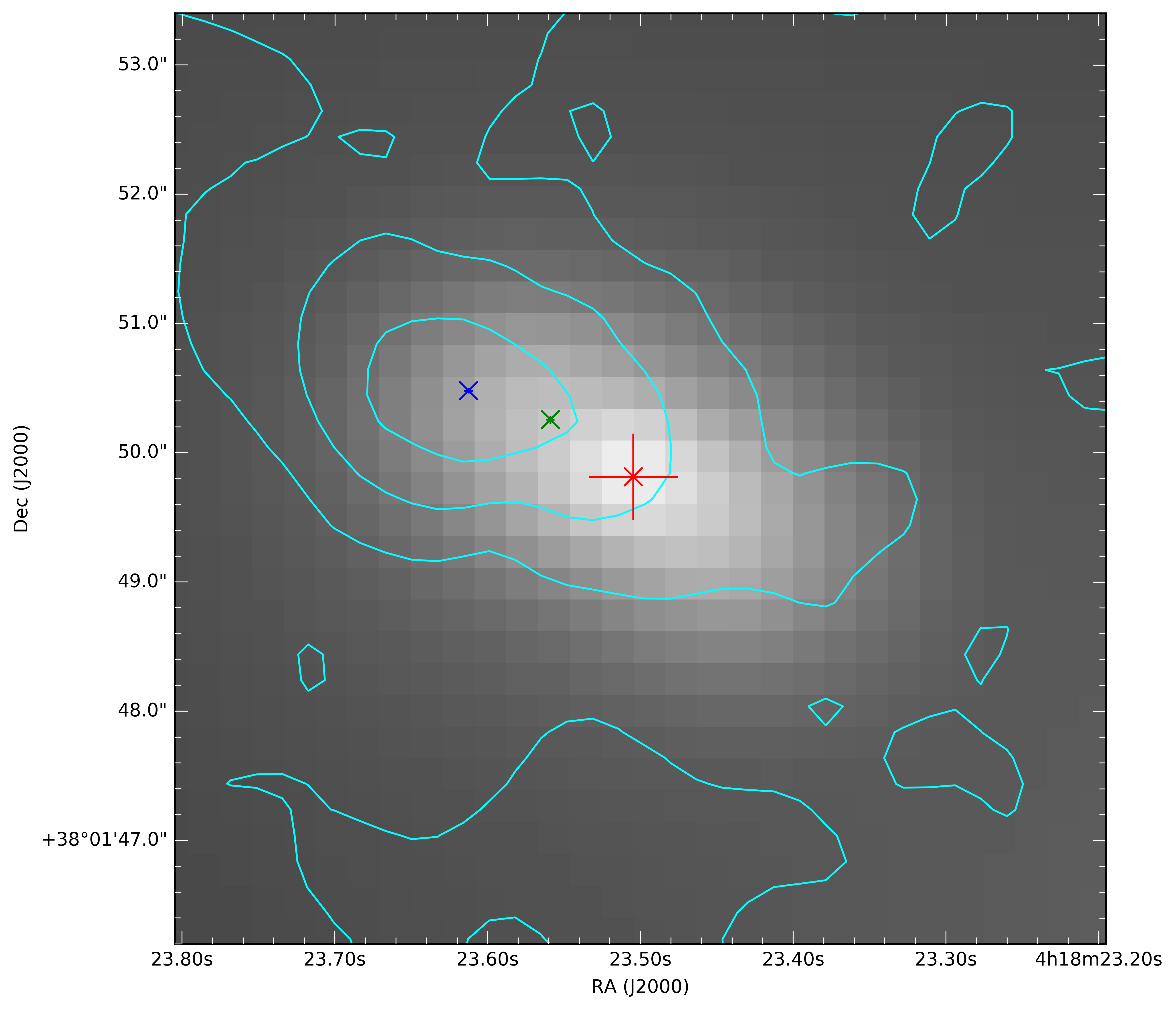}
\includegraphics[width=8.5cm]{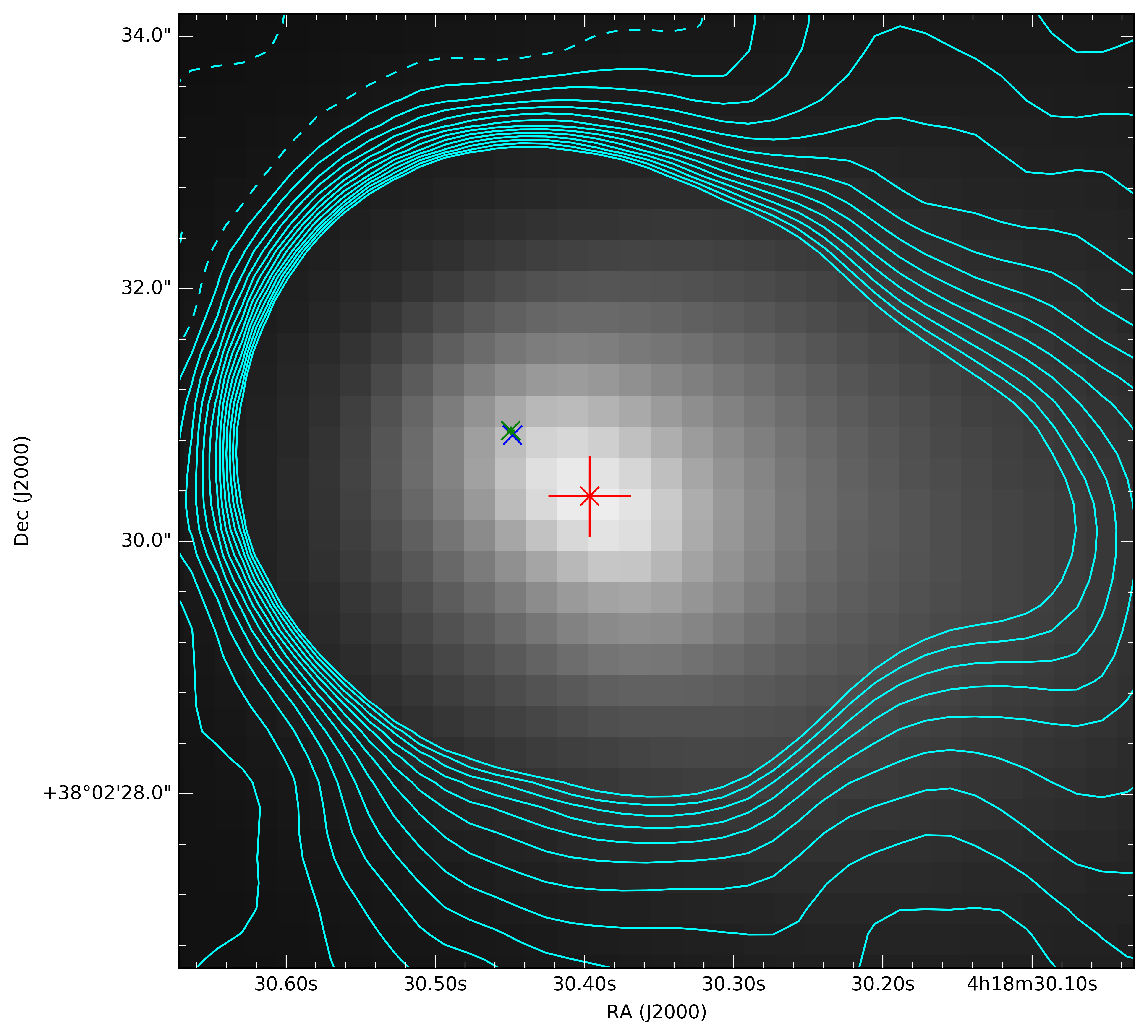}

\caption{Close-up views of the K30 (top) and NHS (bottom) regions of the {\sl Chandra} image, showing the location of the flux maxima in the radio (blue), near-IR (green), and X-ray (red) bands. The sizes of the error bars on each position are shown. Radio contours overlaid in cyan.}

\end{figure}

Four regions are seen within the extended lobes.  Within the northern lobe 
we see knots K97 and K107, as well as the flux maximum of the NHS itself.  While these three 
hotspots have outwardly similar morphologies in the X-ray, near-IR and radio, close examination 
reveals important differences.  In particular, it is only in the radio that one appreciates the extent of 
the northern lobe, which extends for over 30\arcsec~ in a 'plume' shape that includes both K97 and K107,
In the X-ray and near-IR, we see only the three hotspots (with K97 barely detected), plus extensions to
the NHS in two directions, the first being back upstream pointing at K107, and the second one pointing
off to the southwest parallel to the flux contours defining the lobe's southern edge.  The latter could indicate material outflowing from the hotspot, similar to what has been postulated for the 3C 273 jet
by \cite{Roser96}, while the general shape of the jet in that region indicates that the jet does bend 
as it enters the lobe.   A close look at the NHS itself also reveals that its flux maximum is not located
at the same position in the radio, near-IR/optical and X-ray, with the X-ray maximum seen upstream
of the maxima seen in the near-IR/optical and radio (which are aligned with each other).  This misalignment, which is suggestive but not firm at the 2.5-3 $\sigma$ significance level, is shown in Figures 1 and 2, and quantified in Fig. 4.
In addition, we also see for the first time X-ray and near-IR emission from the SHS.  The radio and
near-IR emission from the SHS flux are well aligned (Fig. 2), while there simply are not enough photons 
detected in the {\sl Chandra} image to firm up the comparison between its X-ray and optical flux
maximum position, as only $32 \pm 8$ counts are seen from the SHS and the X-ray emission is 
significantly extended.

%\begin{figure}[t]
%\includegraphics[width=9cm]{3C111SED.pdf}
%\caption{Spectral energy distributions for four jet and hotspot regions in the 3C 111 system.  See \S 3 for discussion.}
%\end{figure}

\subsection {Jet Spectral Energy Distribution}

We have extracted fluxes and spectral energy distributions (SEDs) for all jet and hotspot regions.  
The sky regions used are shown in Figure 1 as green ellipses. The results are given in Table 2. Where a component is not detected in a given band, we give a 2$\sigma$ upper limit.  
The optical and near-IR fluxes were 
extracted from the galaxy-subtracted images and are corrected for extinction using the published 
value of $A_V$.   In addition, for the near-IR and optical fluxes we also subtracted the average flux from 
a radial ring at the same distance from the nucleus, in order to minimize galaxy subtraction residuals.  
This was necessary because of the rather disturbed morphology of the host galaxy as well as the 
presence of several bright, nearby companion galaxies as well as bright stars.  To convert the optical 
and near-IR count rates into fluxes we used the header information from SYNPHOT.  By default, these 
assume a flat spectrum ($\alpha \approx 0, F_\nu \propto \nu^{-\alpha}$); however, the errors from 
this assumption are typically $<5\%$ in these wide bands.  The fluxes in a given band are assumed to 
be centered at the band's pivot wavelength.

The X-ray spectra of the three brightest regions in the 3C 111 jet (knots K30 and K61, and the 
NHS) were extracted using {\sl specextract}.   
The extraction regions used are shown on Figure 1.   
Background spectra  
were obtained using annular regions at the same radii as the components itself, and excluding the 
readout streak.   
We used unweighted ARFs and RMFs and corrected for the PSF. 
Spectral fitting was done in {\it Sherpa} using {\it XSpec} models xsphabs and xspowerlaw.

Determining the correct column density of absorption for 3C 111 is complicated, as it is known that the source shows an additional absorbing column in excess of the Galactic value of $N_H=3.0 \times 10^{21} {\rm cm^{-2}}$ (e.g., \cite{1998MNRAS.299..410R, 2011MNRAS.418.2367B, 2013MNRAS.434.2707T}). For this analysis, we have used Galactic absorption with a column density of $N_H = 8.6 \pm 0.02 \times 10^{21} {\rm ~cm^{-2}}$. This was determined from the {\sl Chandra} HETG data set, the full analysis of which will be discussed in a future paper (Tombesi et al., in prep.). The other {\sl Chandra} data sets of 3C 111 suffer from pileup in the region of the quasar nucleus, making it impossible to determine accurately the value of $N_H$ from them $-$ {\it e.g.,} using our 0.3s frame time data to fit the absorption gives a value of $5.04 \times 10^{21} {\rm ~cm^{-2}}$. An $N_H$ of $8.6 \times 10^{21} {\rm ~cm^{-2}}$ is consistent with previous expectations (see also \cite{2013MNRAS.434.2707T}).

We used the CSTAT statistic in Sherpa as well as the Simplex (aka Nelder-Mead) fitting optimization 
method because of their robustness in low-signal cases.  These fits were all checked using the 
Monte-Carlo method, and the results matched those of Simplex. The CSTAT statistic in {\it Sherpa} is equivalent to the Cash statistic but allows for easier checking of the goodness-of-fit. We checked the goodness-of-fit in two ways: first, by looking at the reduced statistic; and second, by running a simulation of the model and using {\sl plot\_cdf} to check that the cumulative distribution function had a median at about 0.5. The fitting was done for 0.5-7 keV on unbinned data. The flux was determined from the {\sl calc\_energy\_flux} function, over a range of 0.5 to 7 keV. Simulations were also used to determine the error in the flux value. Errors in flux are given at 68\% confidence, while the error in photon index and normalization are given at 90\% confidence intervals.
This yielded the X-ray spectral indices given in Table 3.  As can be seen, all three regions have X-ray spectral indices between $\alpha=0.76$ to $\alpha=1.01$.
The X-ray fluxes from other jet regions were corrected for scattered light from the AGN itself using annular regions at the same radius as the component itself. The X-ray count rates for all jet regions were converted to flux assuming Galactic absorption. For the three regions where X-ray spectral fitting was possible, we used the power-law fits given in Table 3. For all other regions, we used a power-law index of $\alpha=0.87$, equal to the average of the three regions fit.

\begin{deluxetable}{crcc}
\tablecolumns{4}
\tabletypesize{\small}
\tablewidth{0pt}
\tablecaption{Jet Component X-ray Spectra}
\tablehead{\colhead{Region} & \colhead{Normalization} &\colhead{$\alpha$} & \colhead{$\chi^2_\nu$} }
\startdata
K30 & $4.27 \pm 0.58 \times 10^{-14}$ & $0.76\pm 0.29$ & 0.944 \\
K61 & $8.28 \pm 0.82 \times 10^{-14}$ & $1.01\pm 0.21$ & 0.985 \\
NHS & $3.49 \pm 0.48 \times 10^{-14}$ & $0.83\pm 0.28$ & 0.927
\enddata
\end{deluxetable}

\begin{figure*}[th]
\includegraphics[width=8.5cm]{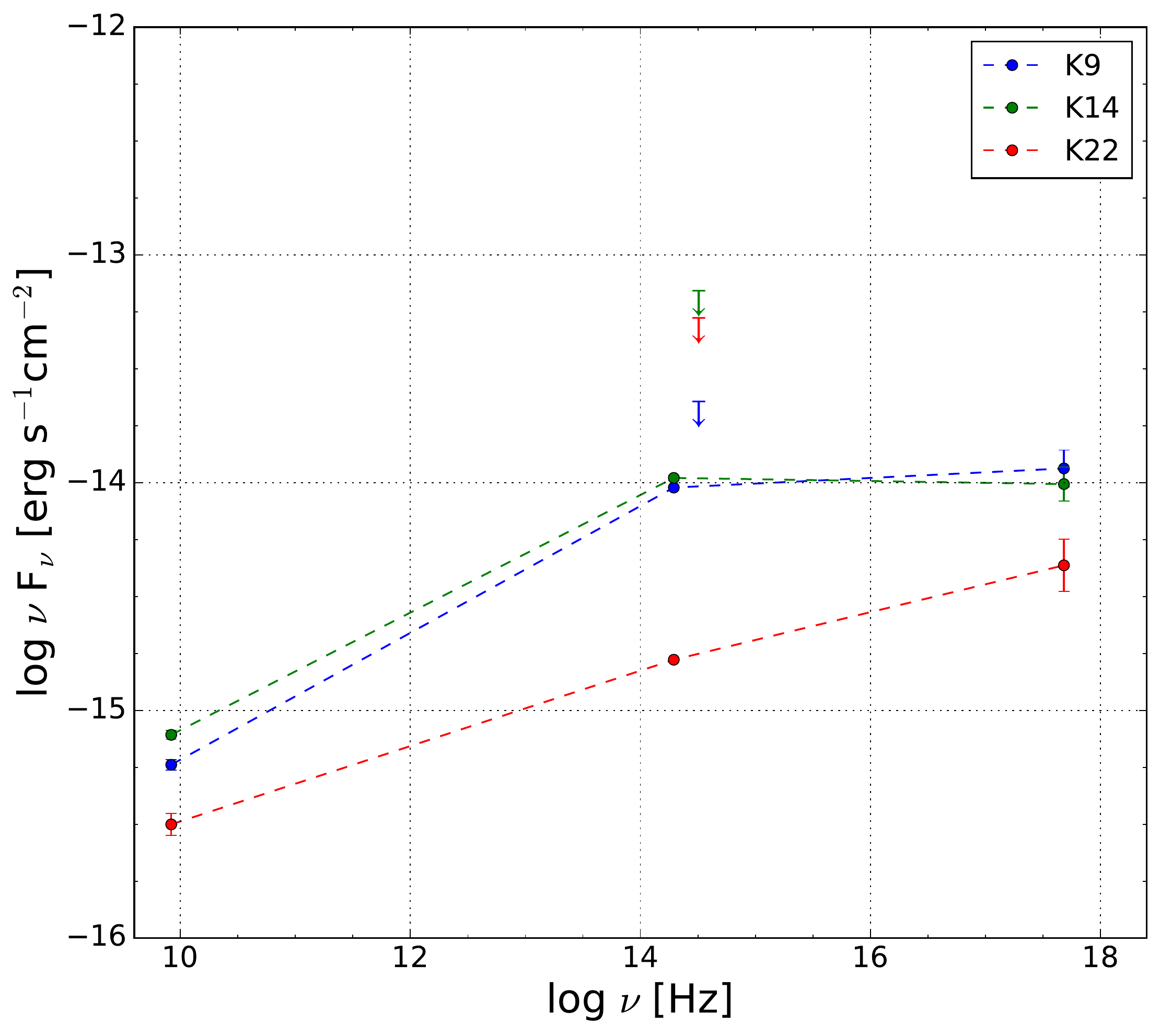}~~~~
\includegraphics[width=8.5cm]{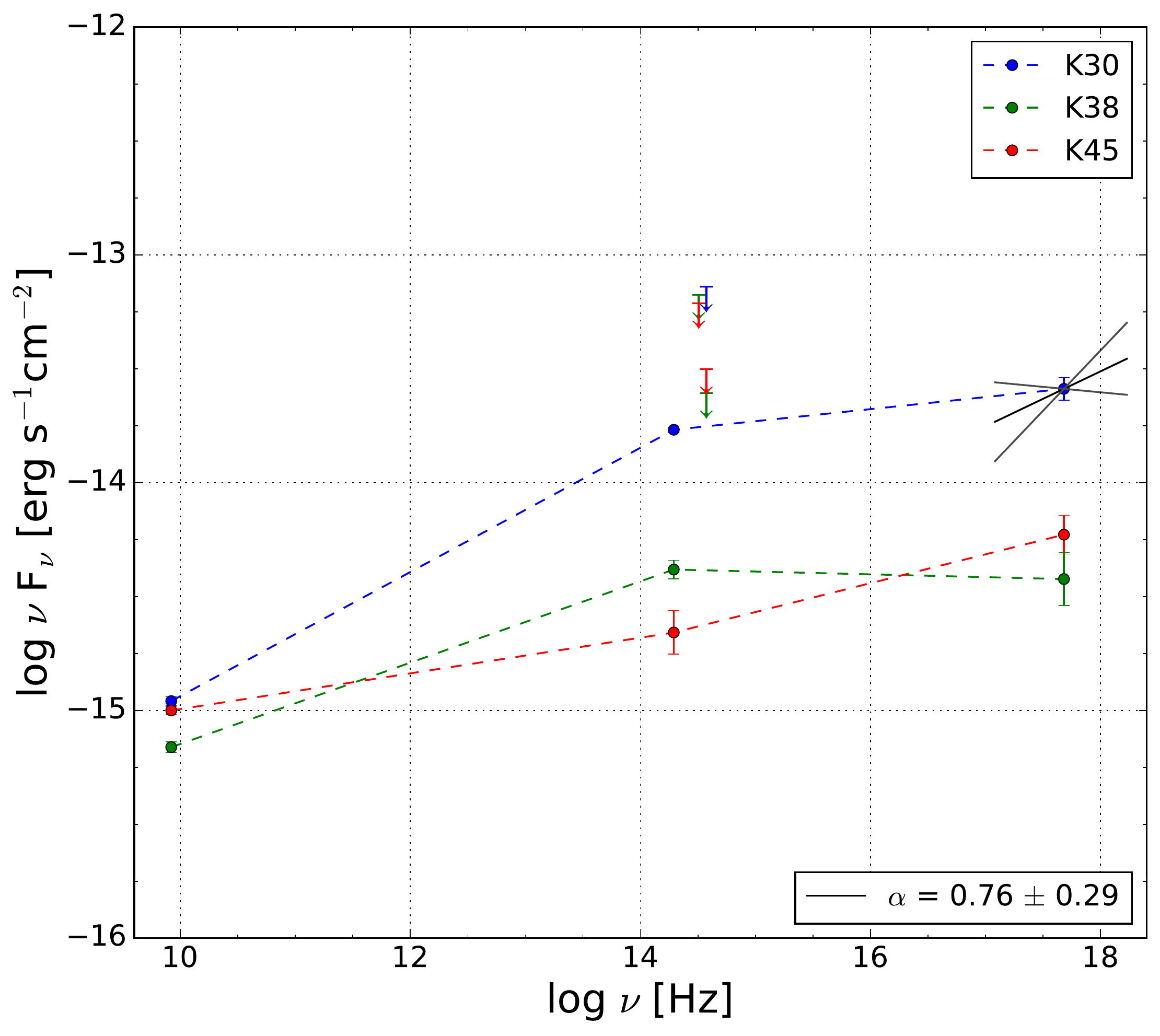} \\
\includegraphics[width=8.5cm]{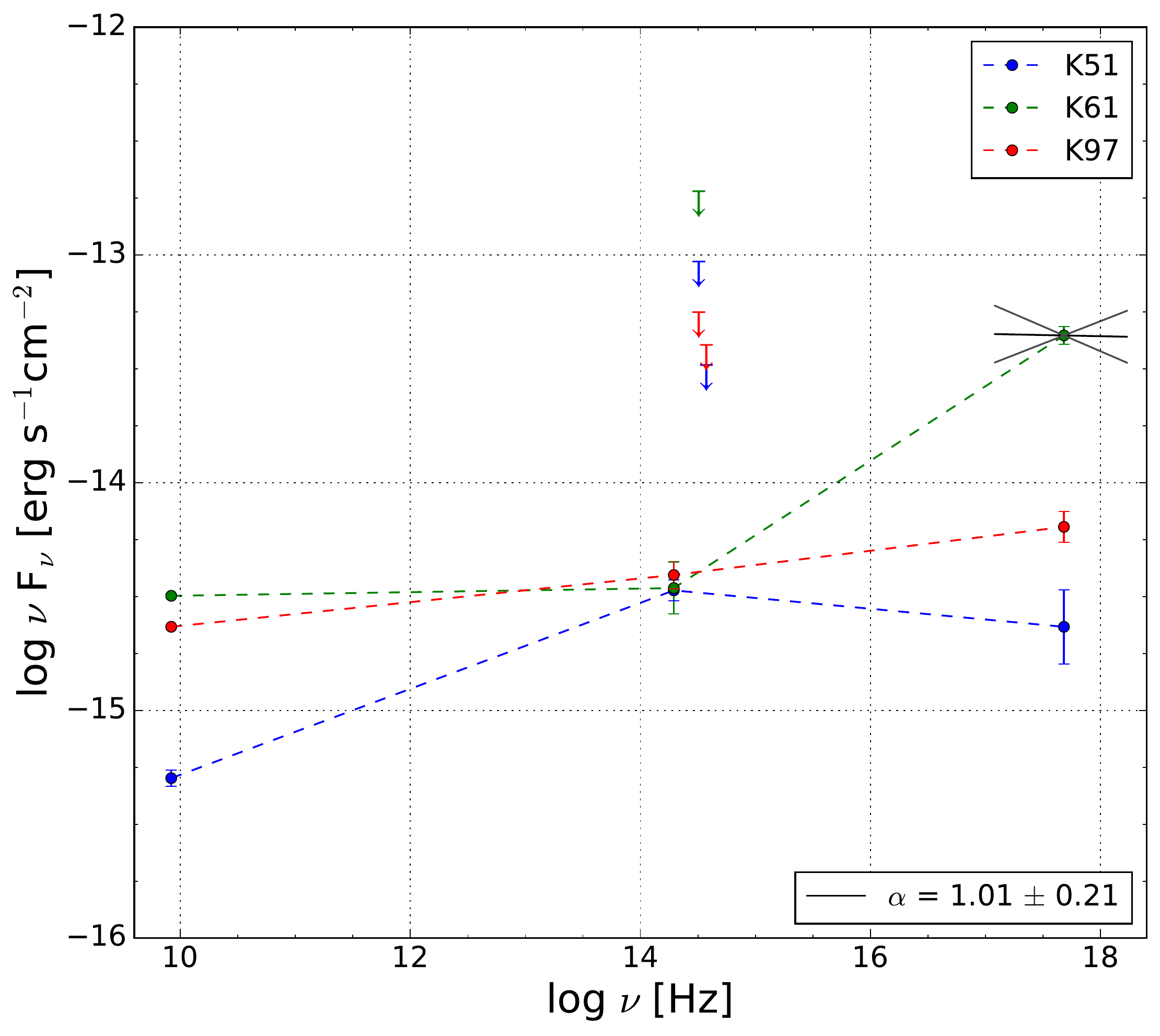}~~~~
\includegraphics[width=8.5cm]{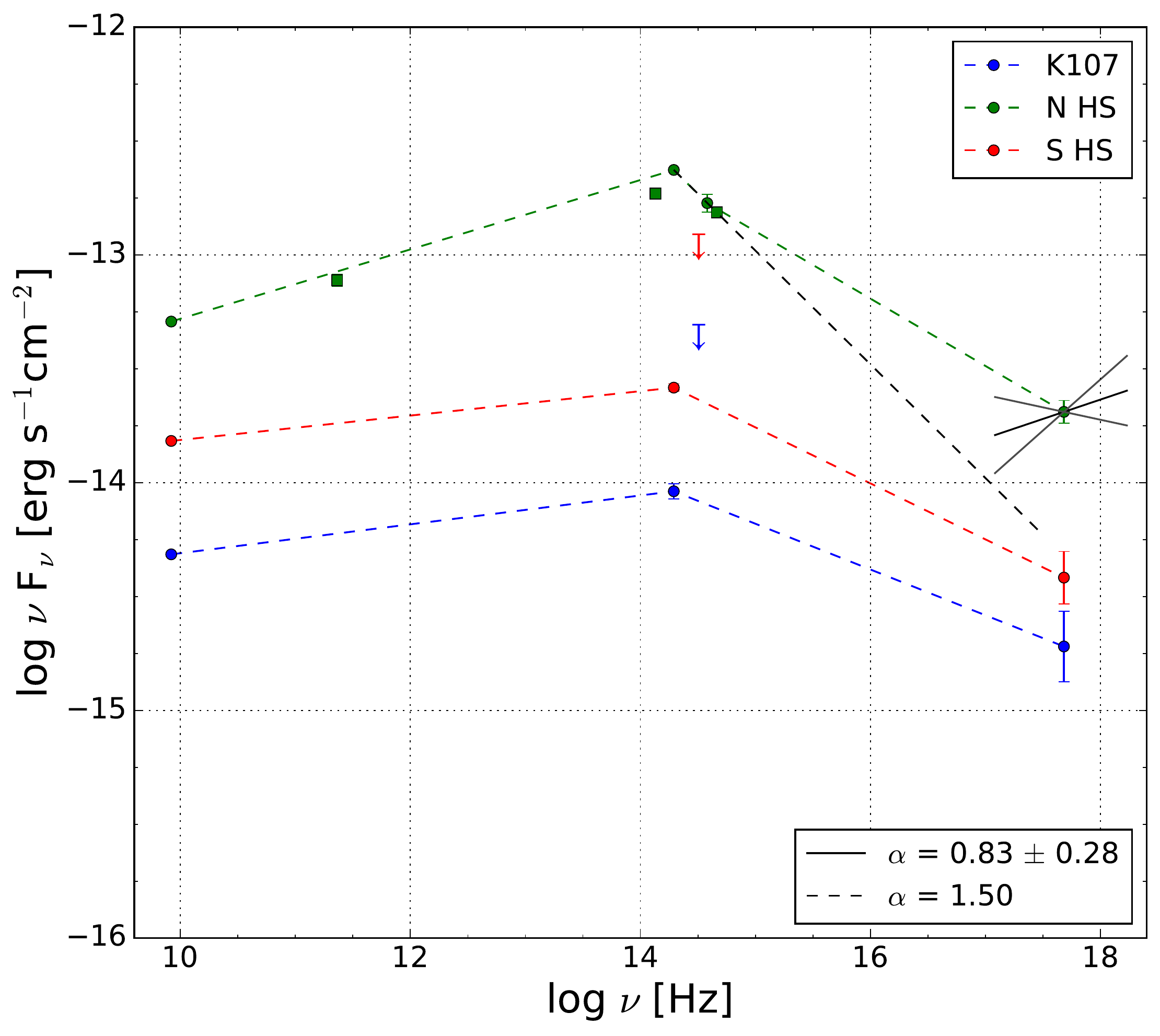}
\caption{Spectral energy distributions for the jet and hotspot regions.   We have
connected all detected fluxes by simple power laws (= straight lines in these log-log plots). Where X-ray spectra are fit, these fits are indicated.  In the NHS, we also plot as the black dotted line the observed near-IR to Optical spectral index.  See \S 3.2 for discussion.}
\label{SED_plot}
\end{figure*}

We show the resulting SEDs for all the components in Figure 5. For regions K30, K61 and the NHS, we use the fitted X-ray flux and spectral index.
Fig. 5 also includes ground-based K and R-band fluxes for the NHS that were 
previously published in \cite{meisenheimer97}, corrected with updated 
values for the Galactic extinction (squares in the lower right panel, see discussion in \S 2.2), as well as a 
1.3 mm flux from IRAM \citep{meisenheimer89}. The 1.3 mm IRAM point lies very close to the power law ($\alpha_R=0.85$) extrapolated from the 8.4 GHz observation of \cite{leahy97}.
The apparent discrepancy between our F160W flux (Table 2, circles in Fig. 5) and the extrapolation
of the K-to-R band spectral index from \cite{meisenheimer97}
%the K and R-band fluxes in \cite{meisenheimer97} and our fluxes (Table 2, circles in Fig 4) 
merits further discussion.  
%One contributor is the larger 
%aperture we used ($5/arcsec$ radius compared to $2.5/arcsec$  in \cite{meisenheimer97}). 
We chose a slightly larger aperture than \cite{meisenheimer97}, to include 
faint extended flux not seen by those authors, as shown in Figure 6. This is only 3\% of the total, and 
both after and before this, our F791W flux is within 1$\sigma$ of the \cite{meisenheimer97} 
K to R band extrapolation.  While it is possible that 
%
%The larger
%aperture was chosen  because of 
%the fact that the NHS has significant extended structure (Figure 3) 
%not seen in the shallower and lower-resolution
%images of \cite{meisenheimer97}.   A smaller, $2.5/arcsec$  aperture (diamonds) decreases the 
%F160W flux by 17\%, still leaving a significant discrepancy, but it
%decreases the F791W flux much more, making it consistent with the R band flux from 
%\cite{meisenheimer97}.  
%Because the implied spectral index with the smaller aperture is so different, we
%are left to consider three possibilities:  
our flux in F160W is incorrect, we consider this unlikely given our careful choice of a  
source-free background region (Figure 6)  and the well-established nature of the HST flux scale.
Alternately, the K-band flux measurement of \cite{meisenheimer97} 
was affected by either poor background subtraction or poor flux calibration.  
We favor this explanation, as due to the crowded field (Figures 3, 6) it is likely 
that the background region in a ground-based image, like that of  \cite{meisenheimer97}, would include flux from one or more neighboring objects, thus causing an apparent underestimate of the source flux.
We were unable to confirm this with the authors of \cite{meisenheimer97}, however).

\begin{figure}
\includegraphics[width=8.6cm]{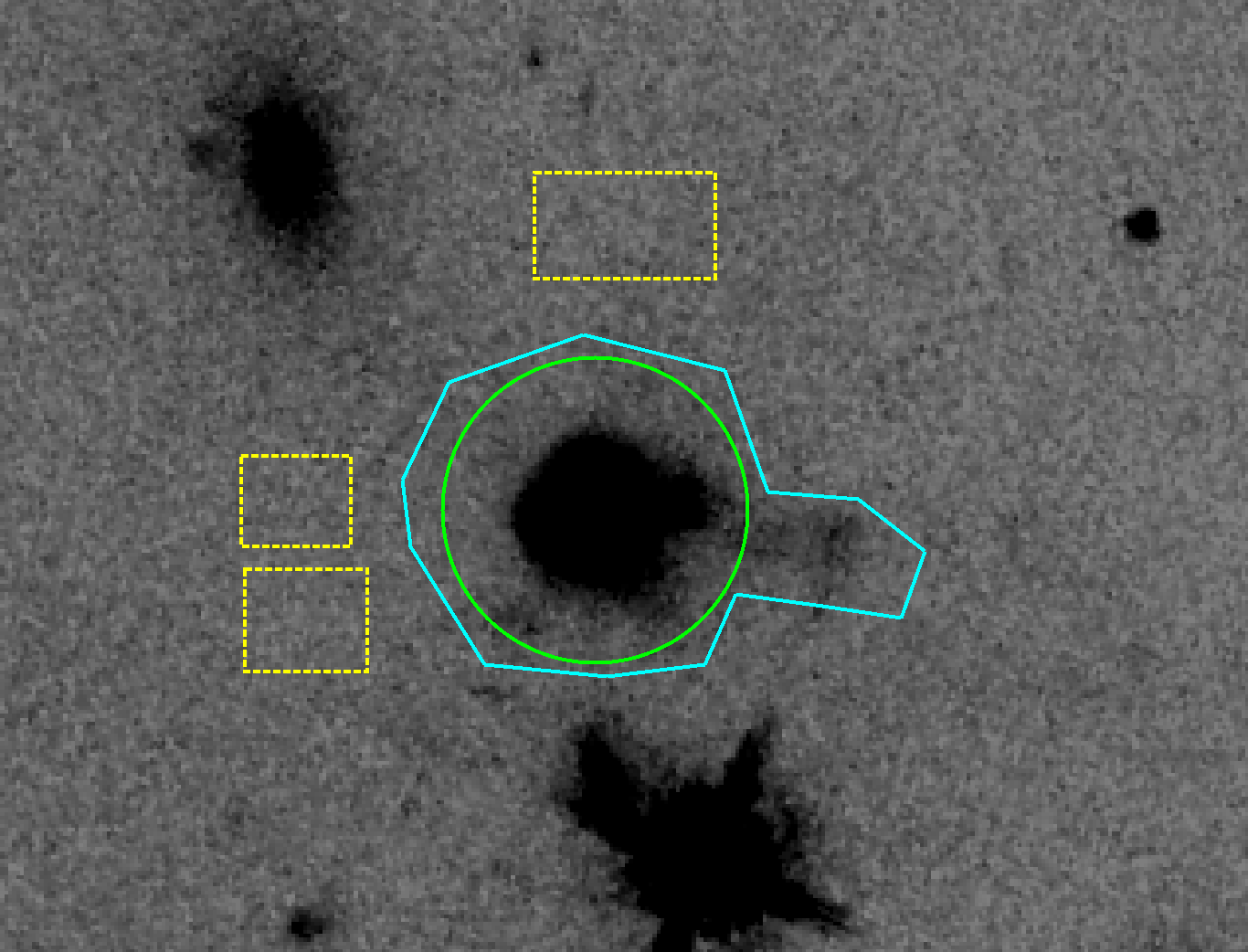}
\caption{Close-up of the NHS region of the F160W image.  The smaller region used by Meisenheimer et al. (believed to be
centered on the NHS) is shown in 
green, while the larger region we use is shown in light blue.  Our background regions are the 
yellow rectangles.  As can be seen, the Meisenheimer et al. region 
did not include a small amount of extended flux from the NHS.  
This only makes a small difference in the flux, as we discuss in \S 4.  However, 
the region is quite crowded, making the choice of the background region very sensitive. We believe
that it is likely that the Meisenheimer et al. background region (dimension and location unknown) included some flux from an unrelated foreground
or background source.}
\end{figure}

As can be seen, most of the jet components have 
diverse SED shapes that naively can be fit by synchrotron emission from a single electron
population.  For example, knots K45 and K97 appear to be fit reasonably well by single power laws 
extending up to X-ray energies, and most other knots have X-ray flux that is significantly below 
the extrapolation of the radio-near-IR power law. However, we do not favor this simple 
interpretation, as in the NHS the fitted X-ray spectral slope is much harder than the extrapolation of the radio to optical synchrotron component, while in knot K61 the X-ray flux is a factor of about 4 higher than a simple extrapolation of the radio to near-IR power law. Thus a second emission component is necessary to fit the SED of these jet knots and possibly others.  In broad terms, such a spectral shape has been seen before in other quasar jets \citep[e.g., PKS 0637-752, knots WK7.8 and WK8.7,][]{mehta09}, and requires either a contribution from another, inverse-Compton mechanism (the so-called EC/CMB mechanism), or alternately a second, entirely distinct high-energy electron population to account for the X-ray emission. Here, however, the fitted X-ray spectra combined with the fact that the X-ray emission of knot K30 and the NHS has a maximum at a different location than the near-IR or optical emission, makes the two-component synchrotron interpretation much more likely. Additionally, a Doppler factor of $\delta \gtrsim 45$ is required to explain the observed properties of the NHS flux if EC/CMB is the dominant emission mechanism at work (see \S 4.2).

\section{Physical Considerations}

The jet of 3C 111 is unique for several reasons. Chief among these are the fact that both the approaching and receding hotspots can be seen in all bands, and its extreme length, with X-ray and near-IR components seen in the jet for more than 100 arcseconds. The data we present here can be used to place a variety of constraints on both the kinematics of the jet as well as the X-ray emission mechanism. In \S 4.1, we use the detection of both the approaching and receding hotspots, as well as VLBA observations, to comment on the kinematics of the jet, while in \S 4.2 we discuss efforts to model the X-ray spectrum and broadband spectral energy distribution of the brightest knots to constrain their emission mechanism in the X-rays.

\subsection{Jet Deceleration}

The flux ratio between the northern and southern hotspots can be used to determine the permitted values for $\beta$ and $\theta$ by using 
\begin{equation}
\frac{S_{1}}{S_{2}} = \left( \frac{1 + \beta ~{\rm cos} \theta}{1 - \beta ~{\rm cos}\theta}\right)^{2+\alpha}
\end{equation}

(e.g., \cite{2012rjag.book.....B}).  We do this individually for the radio, near-IR, and X-ray bands.
Here, $\theta$ is the angle to the line of sight, $\beta = v/c$, and $\alpha$ is the spectral index for each band (0.85 for radio, 1.50 for near-IR, and 0.83 for X-ray; see Figure 5, lower right panel and discussion in \S 3.2). The jet/counterjet hotspot flux ratio differs significantly between bands: $3.34 \pm .01$ in the radio, $9.03 \pm 0.36$ in the near-IR, and $5.34 \pm 1.61$ in the X-ray. This equation makes the assumptions that the jet and counterjet are exactly identical and 180$^{\circ}$ apart.
\cite{jorstad2005} used VLBA observations and determined the most likely viewing angle to be $18.1 \pm 5.0$ degrees on the parsec scale. We found the permitted range of $\beta$ and $\theta$ for the VLBA scale by using their value for the transverse $\beta_T$ apparent to solve

\begin{equation}
\beta_T = \frac{\beta ~{\rm sin}\theta}{1 - \beta ~{\rm cos} \theta}.
\end{equation}

\begin{figure}[h]
\begin{center}
\includegraphics[width=8.6cm]{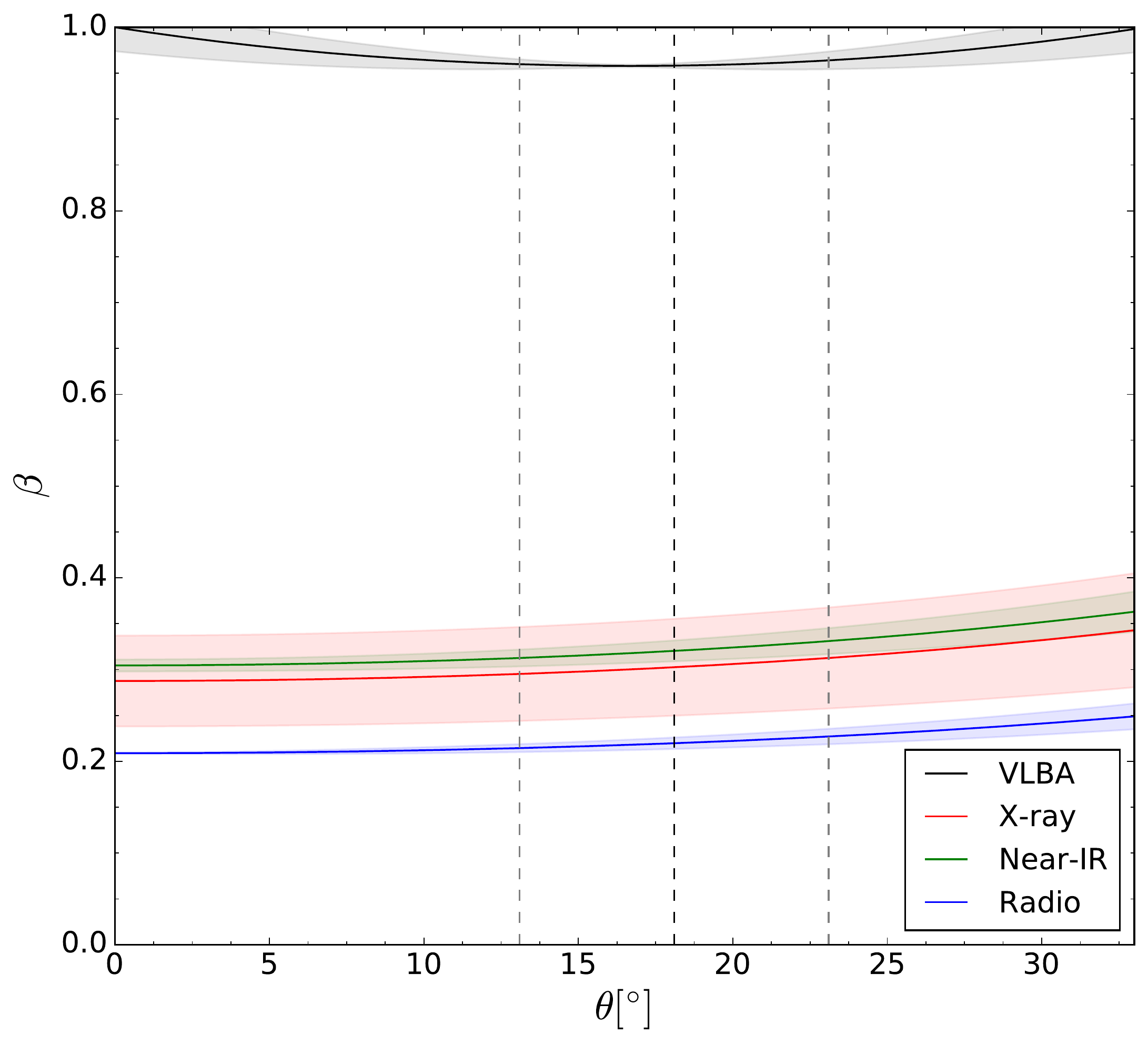}
\caption{Plot of $\beta = v/c$ vs viewing angle for the VLBA scale (solid black line) and kpc-scale radio (blue), near-IR (green), and X-ray (red). 1$\sigma$ uncertainties shown as shaded regions. The dotted lines indicate the VLBA-scale likely viewing angle of $18.1 \pm 5.0$ degrees.}
\label{beta_plot}
\end{center}
\end{figure}

Figure~\ref{beta_plot} shows the $\beta$ vs $\theta$ plot for the parsec-scale VLBA results as well as the $\sim$100 kiloparsec-scale hotspots using our data. We see a clear deceleration from $\beta \sim$ 0.96 at the parsec scale to $\beta \sim$ 0.2-0.4 at the hotspot, with the velocity of the radio-emitting plasma significantly slower than that of the X-ray- and near-IR-emitting plasma. This is consistent with the two-component synchrotron model due to the fact that the radio- and X-ray-emitting electron populations appear to be moving at significantly different velocities, however it may require that the near-IR-emitting electrons do not occupy the entire 
jet cross-section, as in the simplest version of this scenario the near-IR and radio emission come from the same spectral component. Given the relatively modest beaming we find, it is interesting that no jet components are seen in the counterjet between the nucleus and SHS. Additional {\sl HST} and {\sl Chandra} observations are required to better constrain the near-IR spectral index and elaborate on these issues. \cite{2015JKAS...48..299O} more recently used VLBI observations to constrain the viewing angle of 3C 111 on mas scales to $\theta \lesssim 20$ degrees and the intrinsic velocity to $\beta \gtrsim 0.98$, in agreement with the findings of \cite{jorstad2005}. Given the large assumptions and the probable complex structure and dynamics of the hotspot regions, this analysis serves to place an upper limit on the amount of beaming in the jet. The analysis is inconsistent with a highly-beamed jet, as we would expect the jet/counterjet hotspot flux ratio to be larger if beaming were higher.

The spectral index used for the radio is based on the assumption that the slope is constant up to the near-IR. We plan to improve on this value in a future paper where we analyze JVLA observations (C, X, and Ku bands) of 3C 111. A harder spectral index for the radio would increase the likely value for $\beta$, however the offset would not be large enough to bring it into agreement with the near-IR, where the $\Delta \beta \sim 0.1$. This uncertainty does not affect the small $\Delta \beta$ between the X-ray and near-IR, though the near-IR spectral slope could change a small amount with additional {\sl HST} bands to fit the slope.

While the viewing angle has a rather large uncertainty, the $\beta$ value is much more constrained. The relative difference in $\beta$ between bands is preserved no matter the viewing angle, adding to the evidence that there are two electron populations moving at significantly different speeds.

The jet to counterjet length ratio is in relatively good agreement with the radio jet to counterjet flux ratio. The approaching jet is $\sim$ 121 arcsec in length and the counterjet is $\sim 74$ arcsec in length, giving a length ratio of 0.61. For a jet moving at a constant speed $\beta$ and angle $\theta$, we expect the ratio of the lengths to be equal to $(1 - \beta cos \theta) / (1 + \beta cos \theta)$. This matches well with our observed value for $\theta=18.1^\circ$, giving a value of $\beta=0.254$ (Fig. 7), although this depends on how the approaching and receding jets decelerate (e.g., \cite{ryle1967}) and whether there is bending in either jet.

\subsection{Modeling of the Spectral Energy Distribution}

The spectral indices we have obtained for K30, K61, and the NHS are all such that they must lie on either the low-energy tail or near the turnover of the second emission component. Synchrotron and EC/CMB models predict differing slopes for the emission from the very lowest energy electrons, namely $\alpha = -1/3$ for synchrotron and $\alpha = -1$ for EC/CMB (e.g., \cite{dermer2009, stawarz2008}). If the observed spectral index at any part of the low-energy tail were to become significantly harder than $-1/3$, then that would rule out synchrotron as the dominant emission mechanism.

\begin{figure}[ht]
\begin{center}
\includegraphics[width=8.6cm]{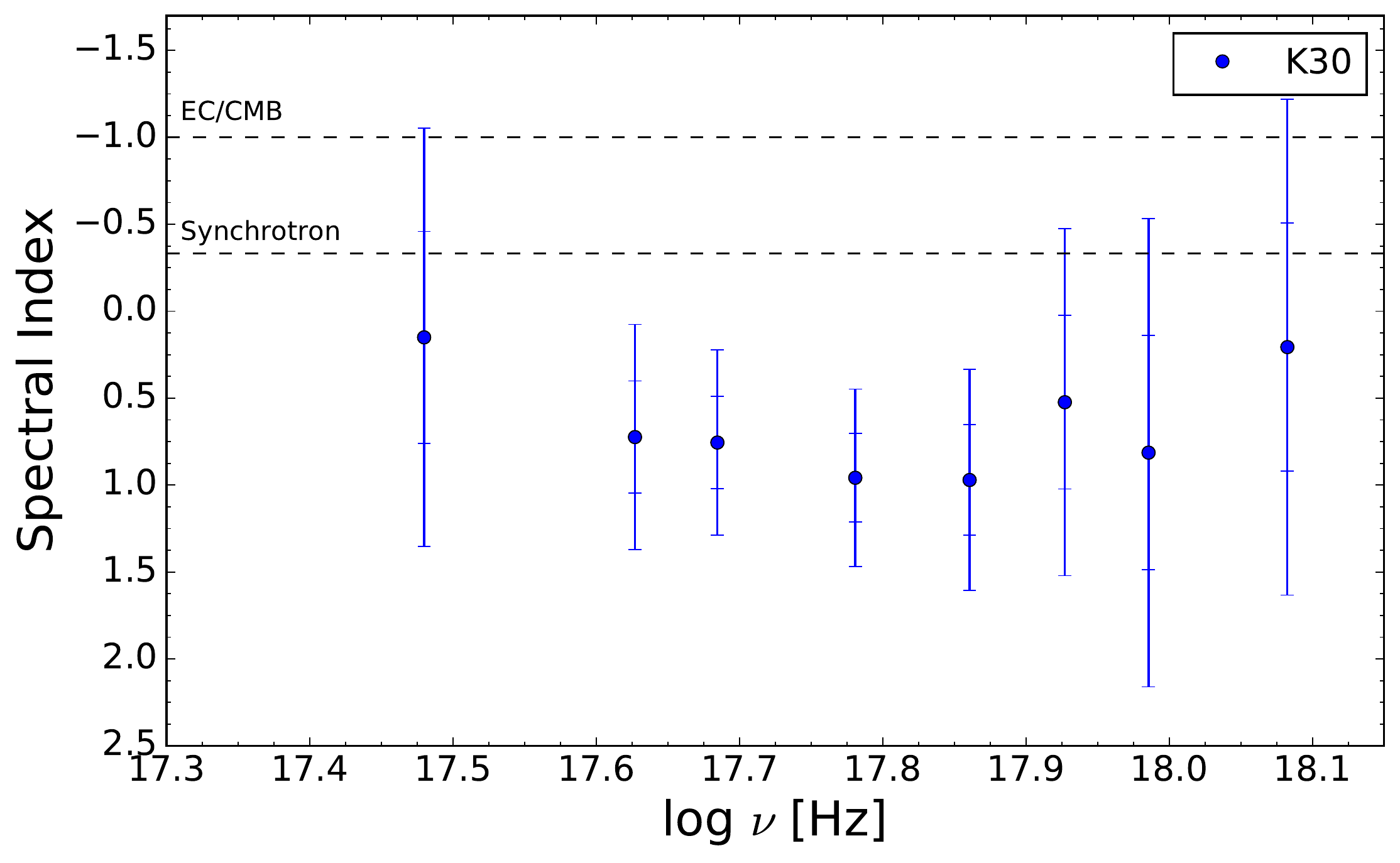}
\includegraphics[width=8.6cm]{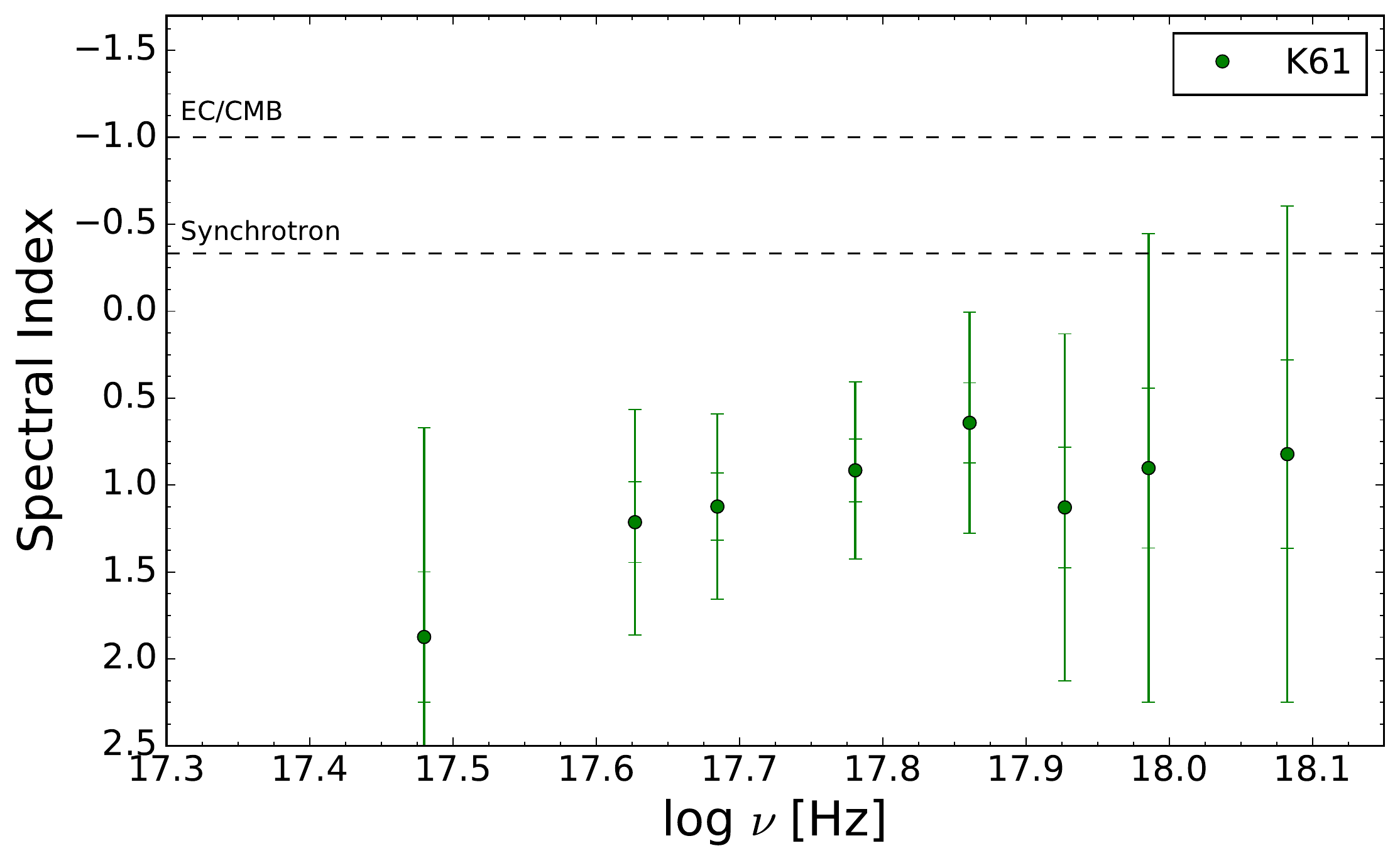}
\includegraphics[width=8.6cm]{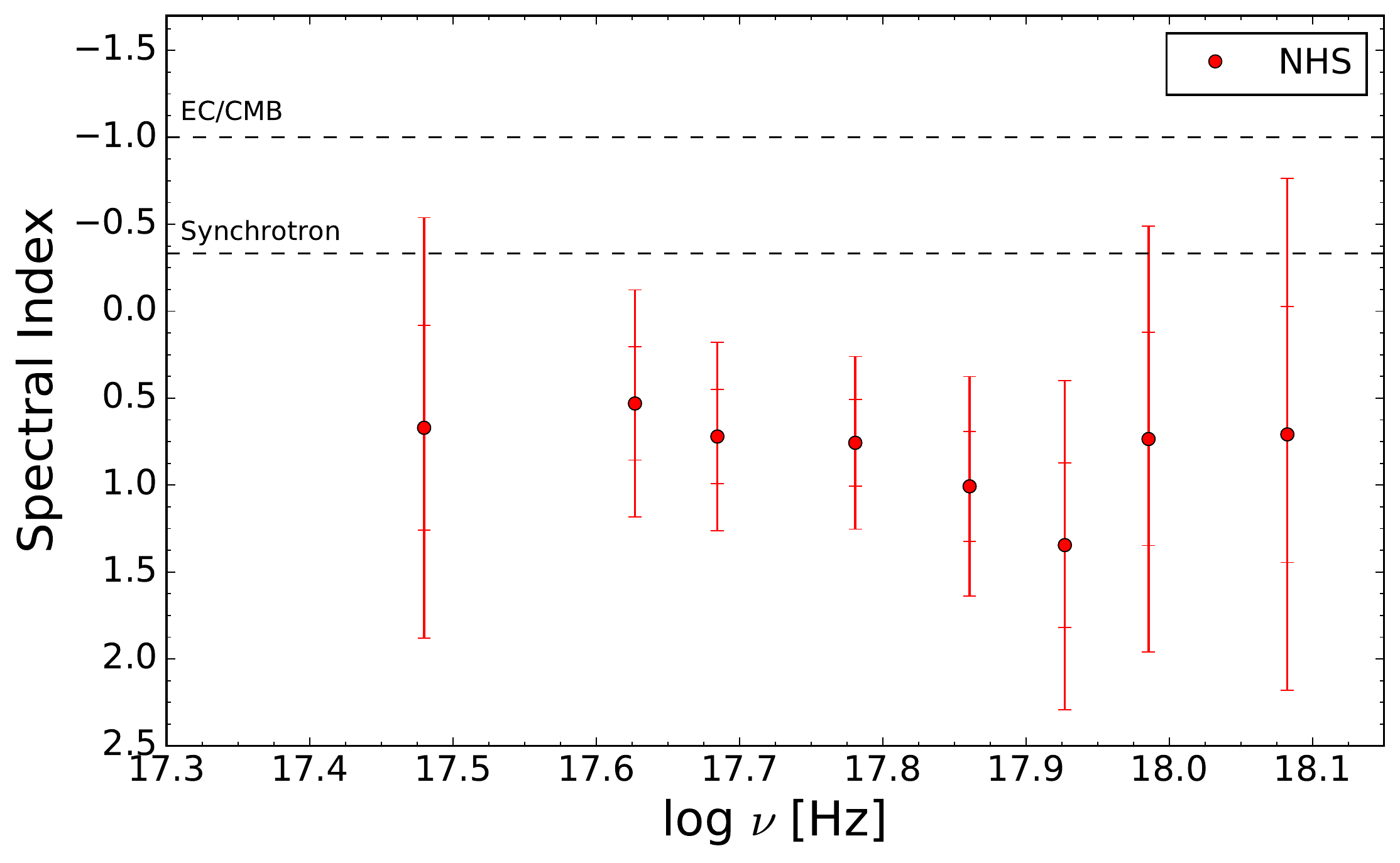}

\caption{X-ray spectral indices for various overlapping energy ranges with error bars for 68\% and 95\% confidence intervals. Energy ranges: 0.5-2, 0.5-3, 0.5-3.5, 1-4, 1.5-4.5, 2-5, 2.5-5.5, 3-7 keV. The labeled dashed lines indicate the predicted spectral indices for the low-energy tail of the synchrotron and EC/CMB models.}
\label{alpha_plot}
\end{center}
\end{figure}

Figure~\ref{alpha_plot} shows the spectral indices for various overlapping energy ranges. All three regions are in good agreement with constant spectral slopes across the entire 0.5-7.0 keV band.

Using the parsec-scale viewing angle of $18.1$ degrees and the associated values for $\beta$ from Figure~\ref{beta_plot}, we can make approximations for the values of $\Gamma$ and $\delta = \lbrack \Gamma (1 - \beta ~{\rm cos} \theta) \rbrack^{-1}$ in order to model the SED for the synchrotron and EC/CMB cases for the NHS.

\begin{figure}[h]
\begin{center}
\includegraphics[width=8.6cm]{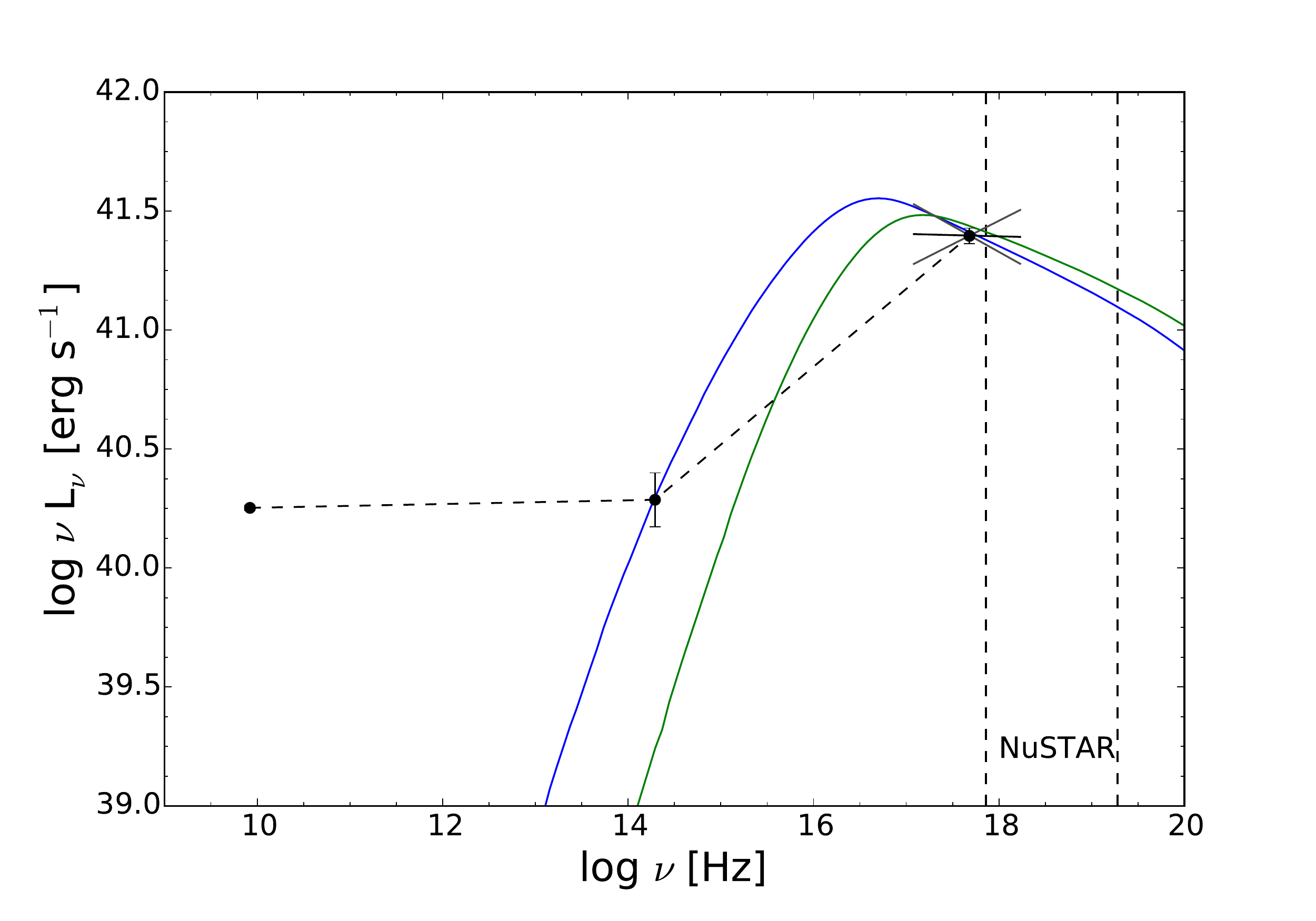}
\includegraphics[width=8.6cm]{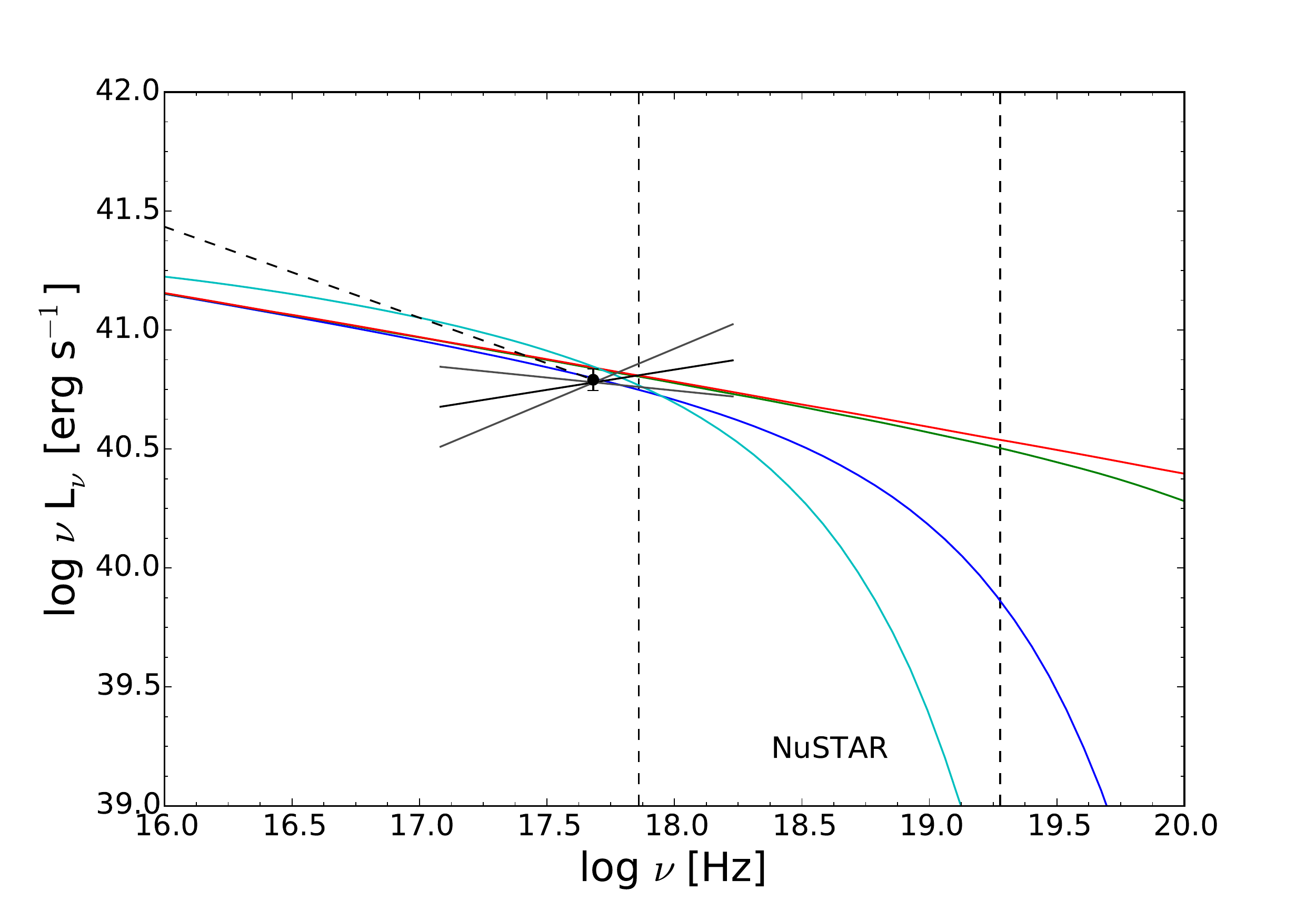}

\caption{Representative models for the SED of the high-energy synchrotron component along with our binned X-ray data for K61 ({\sl Top}) and the NHS ({\sl Bottom}). Dashed vertical lines represent the boundaries of the energy range that {\sl NuSTAR} is capable of observing.}
\label{SED_model}
\end{center}
\end{figure}

Figure~\ref{SED_model} shows several attempts at modeling the SED of K61 and the NHS with varying parameters for the synchrotron model using the Compton Sphere suite{\footnote{Found at http://astro.umbc.edu/compton}}. In the case of K61, our near-IR and X-ray data serve to constrain the low-energy tail of the second emission component. However, because the near-IR spectrum for K61 is not available, we are not able to determine whether the detected flux is dominated by the first or second emission components $-$ the spectrum could be either falling in the near-IR as the first synchrotron component dies off, or it could be rising as the second emission component ramps up. Future {\sl HST} observations would allow us to constrain which emission component is responsible for the detected near-IR flux. We have plotted two example models for the second emission component showing these possibilities using a magnetic field strength ranging over $B = (1-3.2) \times 10^{-5}$ G, with $\gamma_{max} = (3.6-10) \times 10^{9}$, and $\gamma_{min} = (1.3-3.6) \times 10^7$, with a comoving luminosity of $2.15 \times 10^{42}\ {\rm erg\ s^{-1}}$. The magnetic field strength $B$ and fitted $\gamma_{max}$ values translate to a radiative lifetime of $\sim$ 100 years, which is difficult to explain without distributed {\it in situ} acceleration $-$ this requirement can be relaxed by using a lower value of $B$.

Varying several of the input parameters can have a large effect on the shape of the curve above 7 keV for K61 and especially in the case of the NHS. The bottom of Figure~\ref{SED_model} shows several representative models for the SED of the NHS near the {\sl NuSTAR} energy band. Unlike K61, the low-energy tail of the second emission component of the NHS is not constrained by the radio or near-IR data. The models shown here vary wildly in emission above 7 keV, where the magnetic field strengths ranges over $B = (0.2-1) \times 10^{-4}$ G, with $\gamma_{max} = (1.9-100) \times 10^{8}$, and $\gamma_{min} = (5.2-27) \times 10^3$, with a comoving luminosity of $1 \times 10^{43}\ {\rm erg\ s^{-1}}$. Future observations using {\sl NuSTAR} would allow us to constrain the SED up to $\sim$ 80 keV.

% $1 \times 10^{-5}$ G $\leq$ B $\leq 3.2 \times 10^{-5}$ G, with $3.6 \times 10^9 \leq \gamma_{max} \leq 1 \times 10^{10}$, and $1.3 \times 10^7 \leq \gamma_{min} \leq 3.6 \times 10^7$ for a comoving luminosity of .

If the X-ray emission is due only to EC/CMB, then an estimate of the magnetic field strength can be made using

\begin{equation}
\frac{S_{sync}}{S_{IC}} = \frac{(2 \times 10^4 T)^{(3-p)/2} {B_{\mu G}^{(1+p)/2}}}{8 \pi \rho}
\end{equation}

\citep{felten1966}, where $\rho = \Gamma^2 \rho_0 (1 + z)^4$ is the apparent energy density of the CMB at redshift z, $\rho_0 = 4.19 \times 10^{-13}$ erg cm$^{-3}$ is the local CMB energy density, the apparent temperature of the CMB is $\delta T$, and the temperature of the CMB is $T = 2.728(1 + z) K$. This calculation gives a value of $B \approx 7.9 \times 10^{-5}$ G.  While this is comparable to that quoted for other jets where the EC/CMB model is used to model their X-ray emission, in this case a comoving luminosity of $\sim 10^{51} {\rm erg\ s^{-1}} $ is required to fit the model to our X-ray data. We feel this is unrealistic, as it would violate the Eddington limit by many orders of magnitude. For that reason, we have not shown it in any figure.

Additionally, assuming an equipartition magnetic field, a Doppler factor of $\delta \sim 45$ is required for EC/CMB to explain the observed X-ray/radio NHS flux even for the case of $\theta = 0$ degrees using standard formulae \citep{2002ApJ...565..244H}. The required beaming is highly unlikely given the observed properties of the 3C 111 jet, {\it e.g.} the observed brightness of the SHS and the lack of obvious blazar properties.

We do not have many data points with which to constrain the model of the low-energy synchrotron component, especially in K30 and K61. We expect to be able model its SED well in a follow-up paper using JVLA observations of the jet. As well, additional 
{\sl HST} and {\sl Chandra} observations would help to better constrain the near-IR to optical and X-ray spectral indices of the components, and perhaps also constrain the X-ray emission mechanism of additional components.

\section{Conclusions}

We have presented new {\sl Chandra} and {\sl HST} observations of 3C 111 that reveal that its jet has eight X-ray and near-IR/optical emitting components, which extend for 121 arcsec (355 kpc deprojected length) from its AGN nucleus in the approaching jet, and also
reveal the hotspot emission on the counterjet side. The 3C 111 jet is remarkable for several reasons.
While some other jets are comparably long, no other known jet boasts the same combination of length, number of visible components and low redshift that 3C 111 does. For example, the jet of Pic A \citep{marshall10, gentry15, hardcastle15}, which is similarly straight, longer in angular extent (almost 4'), and is about 30\% nearer, has only three components that have been detected in the near-IR, while the jet of 3C 273 \citep{Jester06}, which extends for a somewhat greater distance from its host galaxy and is somewhat brighter, is nearly $4\times$ as far at a redshift $z=0.158$.

The analysis discussed in this paper strongly disfavors the EC/CMB model as the dominant X-ray emission mechanism in several of the components of 3C 111's jet. The hotspot flux ratio for each of the bands we have shows the jet to have decelerated to, at most, $\beta \sim 0.4$. This, combined with a relatively high viewing angle of $\theta \sim 18.1^\circ$ based on VLBA observations, demands a power requirement many orders of magnitude above the Eddington limit for EC/CMB to be the dominant X-ray emission mechanism of the jet.

We instead favor a two-component synchrotron model. Morphological comparison between radio, near-IR, and X-ray bands for K30 and the NHS show the X-ray flux maxima to be significantly upstream of the maxima in the radio, suggesting the presence of two separate electron populations with distinct energy distributions in these regions. This evidence is compounded by the analysis of the jet/counterjet hotspot flux ratio for each band, which shows the near-IR- and X-ray-emitting electrons to be moving at a significantly faster velocity than that of the radio-emitting electron population.

We have made efforts to model the spectral energy distribution of the high-energy synchrotron emission and determine how future observations using {\sl NuSTAR} can be used to constrain the emission mechanism. Future {\sl HST} and {\sl Chandra} observations will allow us to put further constraints on the spectral energy distribution models for the jet components we have analyzed and test the emission mechanism of additional jet components.

\begin{acknowledgments}

These  results are based  on observations made by the {\it Chandra} 
X-ray Observatory (datasets 702798 and 703007) and {\it Hubble Space Telescope} 
(program 13114), as well as the Very Large Array ({\it VLA}, program AB534). 
EP, DC and FT acknowledge support for this work by the National Aeronautics and Space Administration 
(NASA) through Chandra awards G03-14113A (EP, DC) and G04-15103A (FT) issued by the Chandra X-ray Observatory Center, which is operated by the Smithsonian Astronomical Observatory for and on behalf of the National Aeronautics and Space Administration under contract NAS8-03060.  EP and DC also acknowledge support from HST grant GO-13114.01, which was provided by NASA through a grant from the Space Telescope Science Institute, which is operated by the Association of Universities for Research in Astronomy, Inc., under NASA contract NAS 5-26555.  The National Radio Astronomy Observatory is a facility of the National Science Foundation operated under cooperative agreement by Associated Universities, Inc. This research made use of Astropy, a community-developed core Python package for Astronomy \citep{astropy}, hosted at http://www.astropy.org. This research also made use of APLpy, an open-source plotting package for Python hosted at http://aplpy.github.com.

\end{acknowledgments}

\end{document}